\documentclass[a4paper,11pt]{article}

\usepackage[margin=24mm]{geometry}
\usepackage[T1]{fontenc}
\usepackage{lmodern,microtype}
\usepackage{amsmath,amssymb,mathtools,bm}
\usepackage{booktabs,tabularx,array}
\usepackage{graphicx,float}
\usepackage[dvipsnames]{xcolor}
\usepackage[colorlinks=true,linkcolor=MidnightBlue,citecolor=MidnightBlue,urlcolor=BrickRed]{hyperref}
\usepackage[nameinlink,noabbrev]{cleveref}
\usepackage{authblk}
\allowdisplaybreaks
\numberwithin{equation}{section}

\newcommand{\dd}{\mathrm{d}}
\newcommand{\Mpbh}{M_{\rm PBH}}
\newcommand{\fpbh}{f_{\rm PBH}}
\newcommand{\fadm}{f_{a/\textrm{DM}}}
\newcommand{\ga}{g_{a\gamma\gamma}}
\newcommand{\rSch}{r_{\rm Sch}}
\newcommand{\rta}{r_{\rm ta}}
\newcommand{\rcut}{r_{\rm cut}}
\newcommand{\rdB}{r_{\rm dB}}
\newcommand{\rb}{r_b}
\newcommand{\Resc}{R_{\gamma,{\rm esc}}}
\newcommand{\IN}{\mathcal I_N}
\newcommand{\Rfill}{R_{\rm fill}}
\newcommand{\Rfillzero}{R_{\rm fill}^{(0)}}

\newcommand{\dis}[1]{\begin{equation}\begin{split}#1\end{split}\end{equation}}

\newcommand{\bfrac}[2]{{\left(\frac{#1}{#2} \right)  }}
\newcommand{\eq}[1]{Eq.~(\ref{#1})}

\title{Stimulated Axion Decay around Primordial Black Holes and the Diffuse Radio Background}
\author[1]{Ki-Young Choi}
\author[1]{Junghoon Joh}
\author[2]{Kenji Kadota}

\affil[1]{Department of Physics and Institute of Basic Science, Sungkyunkwan University, 2066 Seobu-ro, Suwon-si, Gyeonggi-do, 16419, Korea}
\affil[2]{School of Fundamental Physics and Mathematical Sciences, Hangzhou Institute for Advanced Study, University of Chinese Academy of Sciences (HIAS-UCAS), Hangzhou, 310 024, China}

\date{}

\begin{document}
\maketitle

%%%%%%%%%%%%%%%%%%%%%%%%%%
\begin{abstract}
%%%%%%%%%%%%%%%%%%%%%%%%%%
We study stimulated axion decay in dense axion halos around primordial black holes (PBHs), assuming that both PBHs and axions contribute to dark matter.
We construct the axion density profile by matching a hydrogenic soliton-like core dominated by the PBH potential to the $r^{-9/4}$ spike produced by secondary infall. 
We also impose an inner cutoff at the innermost stable circular orbit.  
We derive local Boltzmann equations for the number densities of photons and axions, including spontaneous and stimulated decay, inverse decay, photon escape, and axion filling.  
We solve these equations numerically and with a semi-analytic approximation. 
Stimulated emission occurs within an inner region where photon amplification overcomes escape.  
We calculate the photon emission from an individual PBH halo and integrate it over the population of PBHs throughout cosmic history. 
By requiring the resulting diffuse radio spectrum not to exceed the radio background measured by ARCADE~2, we derive upper bounds on the fraction of dark matter in PBHs.  
The bounds are strongest at intermediate PBH masses and apply over a wider mass range as the coupling between axions and photons increases.
\end{abstract}

\clearpage
\tableofcontents

%%%%%%%%%%%%%%%%%%%%%%%%%%
\section{Introduction}
%%%%%%%%%%%%%%%%%%%%%%%%%%

The nature of dark matter (DM) remains one of the central open problems in particle physics and cosmology. 
Among the many proposed candidates, axions and axion-like particles are particularly well motivated~\cite{Kim:2008hd,Marsh:2015xka}.
The QCD axion arises from the Peccei--Quinn solution to the strong CP problem~\cite{Peccei:1977hh,Weinberg:1977ma,Wilczek:1977pj} and is expected to be very light and weakly coupled~\cite{Kim:1979if,Shifman:1979if,Zhitnitsky:1980tq,Dine:1981rt}, while light pseudoscalars with similar couplings commonly arise in theories beyond the Standard Model~\cite{Jaeckel:2010ni,Svrcek:2006yi,Arvanitaki:2009fg}. 
Axions can be produced nonthermally in the early Universe through mechanisms such as vacuum misalignment~\cite{Preskill:1982cy,Abbott:1982af,Dine:1982ah} and the decay of topological defects~\cite{Davis:1986xc,Hiramatsu:2012sc}. 
They subsequently behave as cold dark matter and can account for the observed DM abundance~\cite{Kim:2008hd}. 
In this work, we use the word axion for both the QCD axion and a generic axion-like particle with mass $m_a$ and a two-photon coupling $\ga$.

The axion--photon interaction is described by
\begin{equation}
    \mathcal{L}_{a\gamma\gamma}
    = -\frac{1}{4}\ga a F_{\mu\nu}\widetilde{F}^{\mu\nu},
\end{equation}
where $F_{\mu\nu}$ is the electromagnetic field-strength tensor and $\widetilde{F}^{\mu\nu}$ is its dual.  
This interaction allows an axion to decay into two photons with energies near $m_a/2$.  
For the parameters of interest, however, the vacuum lifetime $\tau_a=64\pi/(\ga^2m_a^3)$ far exceeds the age of the Universe, making the spontaneous decay signal extremely weak.  
The bosonic nature of photons leads to Bose enhancement of the decay rate when the final photon states are already occupied~\cite{Tkachev:1987cd,Kephart:1994uy}. 
Such amplification has been studied in idealized homogeneous axion backgrounds~\cite{Arza:2018dcy}, spatially localized clumps and axion stars~\cite{Hertzberg:2018zte,Levkov:2020txo}, and superradiant axion clouds around rotating black holes~\cite{Rosa:2017ury}.  
Related work has further translated stimulated emission from superradiant axion clouds around Kerr primordial black holes into an extragalactic photon flux and corresponding constraints on the PBH abundance~\cite{Garnica:2025qpx}.
In a homogeneous expanding Universe, however, the resonance is suppressed by cosmological redshifting and the photon plasma mass~\cite{Alonso-Alvarez:2019ssa}.  
Efficient stimulated decay therefore requires a gravitationally bound region with high axion density, low velocity dispersion, and a photon residence time long enough for amplification.

Among such gravitationally bound environments, primordial black holes (PBHs) provide a natural setting for dense axion configurations.  
PBHs can form from the collapse of sufficiently large primordial density fluctuations~\cite{Carr:1974nx,Carr:1975qj} and have long been considered as DM candidates~\cite{Carr:1975qj,Chapline:1975ojl,Carr:2020xqk,Green:2020jor}.
A wide range of observations strongly constrains the PBH abundance across much of the PBH mass range~\cite{Green:2020jor,Carr:2020gox}.  
This motivates mixed scenarios in which PBHs constitute only a fraction $\fpbh$ of the DM, while the remainder consists of particle dark matter~\cite{Lacki:2010zf,Boucenna:2017ghj,Carr:2020mqm}. 
Here $\fpbh\equiv\Omega_{{\rm PBH},0}/\Omega_{{\rm DM},0}$ is the present fraction of DM in PBHs.
In such scenarios, each PBH accretes the surrounding particle DM during radiation and matter domination~\cite{Mack:2006gz}. 
Radial infall of cold matter around a point mass produces a steep density spike, $\rho\propto r^{-9/4}$, that extends over many orders of magnitude in radius~\cite{Mack:2006gz,Fillmore:1984wk,Bertschinger:1985pd}. 
For axions, the analogous enhancement appears in the decay channel.  
Close to the PBH, the de Broglie wavelength of the bound axions becomes comparable to the size of the configuration, and the spike is replaced by a PBH-dominated soliton-like core~\cite{Bar:2019pnz,Davies:2019wgi,Aghaie:2023lan}.  
Both the core and the inner spike can reach axion densities far above the cosmological average, making them candidate sites for stimulated axion decay.

Photons produced in PBH halos at redshift $z$ are observed at the redshifted frequency $\nu=m_a/[2h(1+z)]$, where $h$ is Planck's constant.
For $m_a\sim10^{-7}$--$10^{-3}\,{\rm eV}$, the integrated emission from the cosmological PBH population therefore contributes to the diffuse radio background at MHz--GHz frequencies.  
This band has been measured by the ARCADE~2 balloon experiment together with lower-frequency surveys, which report a sky-averaged radio intensity, corresponding to the monopole component, in excess of the cosmic microwave background (CMB) and of the expected contribution from extragalactic radio sources~\cite{Fixsen:2009xn,Seiffert:2009xs}.
Irrespective of its origin, this measured radio monopole provides a conservative upper bound on any additional diffuse radio emission.  
The diffuse radio emission produced by stimulated axion decay in the dense axion configurations surrounding PBHs must not exceed this level, which in turn constrains $\fpbh$ for given axion parameters.

In this work, we develop a framework that connects the axion distribution around an individual PBH to constraints from diffuse radio observations.  
We first construct the axion density profile around a PBH by matching a hydrogenic soliton-like core to the spike density profile and imposing an inner cutoff at the innermost stable circular orbit~\cite{Bardeen:1972fi,Kadota:2023wlm}.  
We then reduce the photon and axion Boltzmann equations to local number density equations that include spontaneous and stimulated decay, inverse decay, photon escape, and the refilling of bound axions from the surrounding halo.  
We obtain the resulting equilibrium profiles both numerically and using a semi-analytic approximation. 
Stimulated enhancement occurs within an inner region that extends to an outer radius $r_{\rm enh}$.
We further check that maintaining this equilibrium does not exhaust the cosmological axion reservoir.  
Finally, we integrate the escaping photon flux over the PBH population and cosmic history, compare the resulting diffuse spectrum with the ARCADE~2 measurement, and derive upper bounds on $\fpbh$ as a function of $\Mpbh$, $m_a$, and $\ga$.  

The paper is organized as follows.  
Section~\ref{sec:profile} constructs the axion density profile around a PBH by matching the soliton-like core to the secondary-infall spike and imposing an inner cutoff.
Section~\ref{sec:kinetics} derives the local number density Boltzmann equations and obtains their equilibrium solutions both numerically and semi-analytically.  
Section~\ref{sec:signal} computes the photon escape rate from an individual PBH halo, the cosmological diffuse intensity, and the resulting radio constraint on $\fpbh$.  
We conclude in Section~\ref{sec:conclusions}.

%%%%%%%%%%%%%%%%%%%%%%%%%%
\section{Axion profile around a primordial black hole}
\label{sec:profile}
%%%%%%%%%%%%%%%%%%%%%%%%%%

In this section, we introduce the axion profile around a PBH.
The structure of the profile depends on the axion de Broglie wavelength.
At radii smaller than the de Broglie scale, the wave nature of the axion becomes important, and axions occupy the ground state of the PBH gravitational potential, forming a hydrogenic soliton-like core~\cite{Bar:2019pnz,Aghaie:2023lan}.
At larger radii, axions behave as individual particles, and secondary infall produces a spike profile~\cite{Carr:2020mqm,Ricotti:2009bs,Inman:2019wvr}.
When the de Broglie scale is sufficiently large compared with the inner cutoff $\rcut$, we match the inner soliton-like core to the outer spike using a prescription adapted from Ref.~\cite{Kadota:2023wlm}, yielding the hybrid profile.
Otherwise, the soliton-like region lies inside the cutoff, and the axion halo is described solely by the spike profile, which we refer to as the spike-only profile.

%%%%%%%%%%%%%%%%%%%%%%%%%%
\subsection{Turnaround radius and secondary-infall spike}
%%%%%%%%%%%%%%%%%%%%%%%%%%

For a cold, nonrelativistic axion component, $\rho_a\propto a_{\rm scale}^{-3}$ after the onset of coherent oscillations, where $\rho_a$ is the cosmological axion density and $a_{\rm scale}$ is the scale factor.  
During radiation domination, $a_{\rm scale}\propto t^{1/2}$ and $H=1/(2t)$, hence $\rho_a(t)\propto t^{-3/2}$.
Balancing PBH gravity with cosmological deceleration gives~\cite{Carr:2020mqm,Adamek:2019gns}
\begin{equation}
 \frac{G\Mpbh}{r^2}\sim H^2r,
 \qquad
 \rta^{\rm RD}(t)\simeq(\rSch t^2)^{1/3},
 \label{eq:rta-rd}
\end{equation}
where $G$ is Newton's constant, $\Mpbh$ is the PBH mass, $r$ is the physical distance from the PBH, $H$ is the Hubble expansion rate, and
$\rta^{\rm RD}(t)$ is the turnaround radius at cosmic time $t$ during
radiation domination. 
We denote the Schwarzschild radius by $\rSch\equiv2G\Mpbh$ in natural units.
If the shell now at $r$ turned around at $t_*(r)$, then $t_*\propto r^{3/2}$.  
Its density is comparable to the background density at turnaround, and therefore
\begin{equation}
 \rho_{a,{\rm sp}}(r)\sim\rho_a[t_*(r)]
 \propto t_*^{-3/2}\propto r^{-9/4}.
\end{equation}
Here $\rho_{a,{\rm sp}}(r)$ is the axion density in the spike, $\rho_a[t_*(r)]$ is the cosmological axion density at turnaround, and $t_*(r)$ is the turnaround time of the shell that is located at radius $r$~\cite{Carr:2020mqm,Adamek:2019gns}.
At equality this profile is written as~\cite{Carr:2020mqm,Adamek:2019gns}.
\begin{equation}
 \rho_{a,{\rm sp}}(r)
 =\fadm\rho_{m,{\rm eq}}\left(\frac{r_{\rm eq}}{r}\right)^{9/4},
 \label{eq:spike}
\end{equation}
where $\fadm$ is the axion fraction of the matter density, $\rho_{m,{\rm eq}}$ is the matter density at equality, and $r_{\rm eq}\equiv\rta(t_{\rm eq})$ is the turnaround radius at the matter--radiation equality time $t_{\rm eq}$.

After matter--radiation equality, the PBH acts as a localized excess mass and seeds the gravitational growth of the surrounding matter perturbation.
For a spherical shell enclosing an unperturbed background mass $M$, the initial overdensity is approximately $\delta_{\rm eq}(M)\sim\Mpbh/M$.
During matter domination, the linear density contrast grows as $\delta(M,z)\propto a_{\rm scale}\propto(1+z)^{-1}$.  
A shell turns around and becomes gravitationally bound when its linearly extrapolated density contrast reaches order unity. 
Therefore, the mass scale of the largest shell that has become bound by redshift $z$ is~\cite{Mack:2006gz}
\begin{equation}
 M_h(z)\simeq\Mpbh\frac{1+z_{\rm eq}}{1+z}
 \propto a_{\rm scale},
 \label{eq:secondary-infall-mass}
\end{equation}
where $z_{\rm eq}$ is the matter--radiation equality redshift.
Here $M_h(z)$ denotes the mass scale of the matter halo gravitationally bound to the PBH, excluding the central PBH itself.  
The successive turnaround and accretion of increasingly distant shells is referred to as secondary infall~\cite{Carr:2020mqm,Mack:2006gz,Ricotti:2009bs}.

Since the background matter density decreases as $\rho_m\propto a_{\rm scale}^{-3}$, the turnaround radius of the newly bound shell scales as
\begin{equation}
 r_{\rm ta}^{\rm MD}(z)
 =r_{\rm ta}(t_{\rm eq})
 \left(\frac{1+z_{\rm eq}}{1+z}\right)^{4/3},
\end{equation}
where $ r_{\rm ta}(t_{\rm eq})
 =\left(2G\Mpbh t_{\rm eq}^2\right)^{1/3}$.
Combining $M_h\propto a_{\rm scale}$ with $r_{\rm ta}^{\rm MD}\propto a_{\rm scale}^{4/3}$ gives
\begin{equation}
 M(<r)\propto r^{3/4},
 \qquad
 \rho_{a,{\rm sp}}(r)
 =\frac{1}{4\pi r^2}\frac{\dd M_a(<r)}{\dd r}
 \propto r^{-9/4},
\end{equation}
where the axion mass follows $M_a(<r)\propto\fadm M(<r)$ when axions constitute a constant fraction $\fadm$ of the clustering matter.  
Thus, in the ideal spherical and self-similar limit, matter-era secondary infall extends the spike to progressively larger radii while retaining the $r^{-9/4}$ profile~\cite{Fillmore:1984wk,Bertschinger:1985pd}.

%%%%%%%%%%%%%%%%%%%%%%%%%%
\subsection{PBH-dominated soliton-like inner profile}
%%%%%%%%%%%%%%%%%%%%%%%%%%

We write the real axion field in the nonrelativistic regime as
\begin{equation}
 \phi(\bm x,t)=\frac{1}{\sqrt{8\pi G}}
 \left[e^{-im_a(1+\gamma)t}\psi(\bm x)+{\rm c.c.}\right],
\end{equation}
where $\phi$ is the real axion field, $m_a$ is the axion mass, $\psi$ is its nonrelativistic wavefunction, $\gamma$ is the dimensionless binding-energy eigenvalue, and ${\rm c.c.}$ denotes the complex conjugate.
The spherically symmetric Schr\"odinger--Poisson equations are~\cite{Bar:2019pnz}
\begin{equation}
 (u\psi)''=2u\left(\Phi-\frac{A_G}{u}-\gamma\right)\psi,
 \qquad
 (u\Phi)''=u\psi^2.
 \label{eq:sp}
\end{equation}
Here $u\equiv m_ar$ is the dimensionless radius, $A_G\equiv G\Mpbh m_a$ is the dimensionless PBH gravitational coupling, $\Phi$ is the axion self-gravity potential, and a prime denotes a derivative with respect to $u$.
For $M_{\rm sol}/\Mpbh\ll1$, axion self-gravity is subleading and the ground-state solution is hydrogenic~\cite{Bar:2019pnz,Davies:2019wgi},
\begin{equation}
 \psi(r)=\lambda^2e^{-r/\rdB},
 \qquad
 \rho_{a,{\rm sol}}(r)=\rho_c e^{-2r/\rdB}.
 \label{eq:soliton}
\end{equation}
Here $M_{\rm sol}$ is the soliton-like cloud mass, $\lambda$ fixes the wavefunction normalization, and $\rho_c$ is the central density.
We denote the gravitational de Broglie radius by
$\rdB\equiv\lambda_{\rm dB}=1/(G\Mpbh m_a^2)$; equivalently,
$\lambda_{\rm dB}=1/(m_av_{\rm sol})$ with
$v_{\rm sol}=G\Mpbh m_a$.
This is a PBH-dominated hydrogenic axion core~\cite{Bar:2019pnz,Davies:2019wgi}.
Related formation of axion stars through gravitational condensation in PBH-seeded axion minihalos has been studied in Ref.~\cite{Yin:2024xov}.

%%%%%%%%%%%%%%%%%%%%%%%%%%
\subsection{Construction of the hybrid soliton--spike profile}
%%%%%%%%%%%%%%%%%%%%%%%%%%

For a nonrotating black hole, the Schwarzschild innermost stable circular orbit (ISCO)\footnote{The ISCO is the smallest radius at which a massive test particle can maintain a stable circular orbit around a nonrotating black hole.} is located at $r_{\rm ISCO}=6G\Mpbh=3\rSch$~\cite{Bardeen:1972fi}.
Following Ref.~\cite{Kadota:2023wlm}, we adopt this radius as a phenomenological inner cutoff, $\rcut=3\rSch$, and match the hydrogenic soliton-like inner profile to the outer spike by requiring continuity of the density and equality of the enclosed masses:
\begin{equation}
 \rho_{a,{\rm sol}}(\rb)=\rho_{a,{\rm sp}}(\rb),
 \qquad
 4\pi\int_{\rcut}^{\rb}\rho_{a,{\rm sol}}r^2\dd r
 =
 4\pi\int_{\rcut}^{\rb}\rho_{a,{\rm sp}}r^2\dd r.
 \label{eq:matching-conditions}
\end{equation}
Here, $\rb$ denotes the boundary between the soliton-like and spike density profiles. Defining the dimensionless radii
\begin{equation}
 x_c\equiv\frac{\rcut}{\rdB},
 \qquad
 x_b\equiv\frac{\rb}{\rdB},
\end{equation}
and the dimensionless cumulative-mass function
\begin{equation}
 F(x)\equiv 1-e^{-2x}\left(1+2x+2x^2\right),
\end{equation}
the matching conditions yield
\begin{equation}
 F(x_b)-F(x_c)
 =
 \frac{16}{3}e^{-2x_b}
 \left(x_b^3-x_b^{9/4}x_c^{3/4}\right).
 \label{eq:finite-xb}
\end{equation}
This implicit equation determines the dimensionless matching radius $x_b$ for a given cutoff $x_c$. 
A nontrivial solution with $x_b>x_c$ exists only for $x_c<9/8$; otherwise, we retain the spike-only profile throughout the radial domain.

In the limit $\rcut\ll\rdB$, or equivalently $x_c\to0$, the matching equation in \eq{eq:finite-xb} then reduces to
\begin{equation}
 F(x_b)=\frac{16}{3}e^{-2x_b}x_b^3,
\end{equation}
which gives the solution $x_b=2.02857\ldots$.  
Therefore, the nontrivial matching solution is
\begin{equation}
 \rb\simeq2\rdB.
 \label{eq:rb-semi-analytic}
\end{equation}
The matching conditions in \eq{eq:matching-conditions} determine the central density of the soliton-like profile as
\begin{equation}
 \rho_c\simeq e^4\fadm\rho_{m,{\rm eq}}
 \left(\frac{r_{\rm eq}}{2\rdB}\right)^{9/4}.
 \label{eq:rho-central-semi-analytic}
\end{equation}
Using $M_{\rm sol}=\pi\rho_c\rdB^3$, we obtain the mass fraction
\begin{align}
 \epsilon\equiv\frac{M_{\rm sol}}{\Mpbh}
 &\simeq \frac{\pi e^4}{2^{3/2}}
 \fadm\rho_{m,{\rm eq}}t_{\rm eq}^{3/2}\Mpbh^{-1}m_a^{-3/2}\notag\\
 &\simeq2.42\times10^{-9}
 \left(\frac{\fadm}{1}\right)
 \left(\frac{\Mpbh}{10^{-6}M_\odot}\right)^{-1}
 \left(\frac{m_a}{10^{-6}\,{\rm eV}}\right)^{-3/2}.
 \label{eq:epsilon-semi-analytic}
\end{align}
The PBH-dominated approximation requires $\epsilon\ll1$.

Combining the soliton-like inner profile with the secondary-infall spike yields the matched axion density profile, which we refer to as the hybrid profile. 
This profile is given by
\begin{equation}
 \rho_a(r)=
 \begin{cases}
  \displaystyle
  \fadm\rho_{m,{\rm eq}}
  \left(\frac{r_{\rm eq}}{\rb}\right)^{9/4}
  \exp\left[\frac{2(\rb-r)}{\rdB}\right],
  & \rcut\leq r<\rb,\\[3mm]
  \displaystyle
  \fadm\rho_{m,{\rm eq}}
  \left(\frac{r_{\rm eq}}{r}\right)^{9/4},
  & r\geq\rb.
 \end{cases}
 \label{eq:matched-axion-profile}
\end{equation}

\begin{figure}[t!]
 \centering
 \includegraphics[width=0.82\textwidth]
 {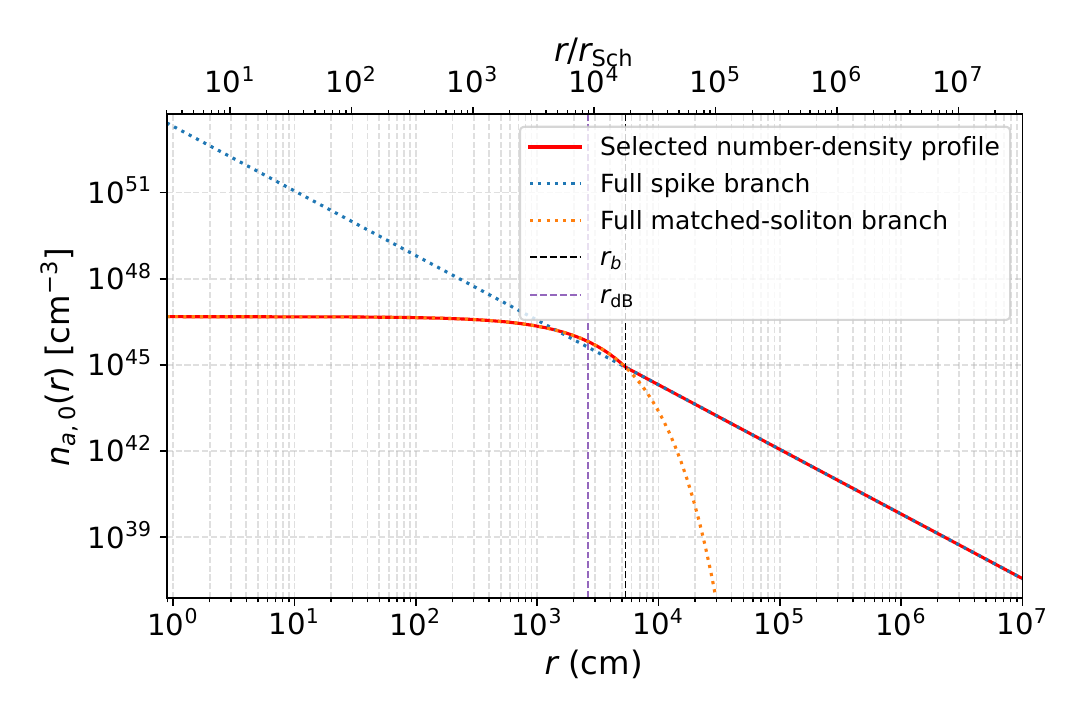}
 \caption{Radial profile of the initial axion number density $n_{a,0}(r)=\rho_a(r)/m_a$.  
 The reference parameters are $M_{\rm PBH}=10^{-6}M_\odot$, $m_a=10^{-6}\,{\rm eV}$, and $\fadm = 1$. 
 The solid curve denotes the hybrid soliton--spike profile, while the dashed red and blue curves show the underlying soliton-like and spike profiles, respectively. 
 The vertical lines mark $r_{\rm cut}=3\rSch$, $r_{\rm dB}=9\times 10^3\rSch$, and $r_b=1.8\times 10^4\rSch$.}
 \label{fig:axion-number-density}
\end{figure}

Fig.~\ref{fig:axion-number-density} shows the radial profile of the initial axion number density $n_{a,0}(r)=\rho_a(r)/m_a$ surrounding the PBH with $M_{\rm PBH}=10^{-6}M_\odot$, $m_a=10^{-6}\,{\rm eV}$, and $\fadm \simeq 1$.
Inside the matching radius $r_b$, the profile is described by the PBH-dominated soliton-like solution, which is constant for $r\ll r_{\rm dB}$. 
At $r_b$, the inner solution is matched continuously to the secondary-infall spike.
Outside this radius, the number density follows the spike profile, $n_{a,0}(r)\propto r^{-9/4}$.
The region $r<r_{\rm cut}=3r_{\rm Sch}$ is excluded from the calculation.

%%%%%%%%%%%%%%%%%%%%%%%%%%
\section{Photon--axion kinetics}
\label{sec:kinetics}
%%%%%%%%%%%%%%%%%%%%%%%%%%

We consider a real axion-like particle with Lagrangian
\begin{equation}
 {\cal L}=-\frac14F_{\mu\nu}F^{\mu\nu}
 +\frac12\partial_\mu a\partial^\mu a-\frac12m_a^2a^2
 -\frac14\ga aF_{\mu\nu}\widetilde F^{\mu\nu},
 \label{eq:lagrangian}
\end{equation}
where $a$ is the axion field, $F_{\mu\nu}$ is the electromagnetic field-strength tensor, $\widetilde F^{\mu\nu}$ is its dual, and $\ga$ is the axion--photon coupling. 
The axion can decay into two photons spontaneously in the vacuum with the decay rate 
\begin{equation}
 \Gamma_a=\frac{\ga^2m_a^3}{64\pi},
 \qquad \tau_a=\Gamma_a^{-1},
 \label{eq:decay-rate}
\end{equation}
with its corresponding lifetime $\tau_a$.
However, in a medium, Bose enhancement leads to stimulated axion decay, while inverse decay can also occur.

%%%%%%%%%%%%%%%%%%%%%%%%%%
\subsection{Boltzmann equations around a PBH}
%%%%%%%%%%%%%%%%%%%%%%%%%%

The general kinetic equation for a phase-space distribution $f_i(\bm x,\bm p,t)$, where $i$ labels the axion and photon species, is
\begin{equation}
 \left(\partial_t+\dot{\bm x}\!\cdot\!\nabla_{\bm x}
 +\dot{\bm p}\!\cdot\!\nabla_{\bm p}\right)f_i=C_i[f],
 \label{eq:general-boltzmann}
\end{equation}
where $C_i[f]$ is its collision operator, and $\dot{\bm x}$ and $\dot{\bm p}$ are the phase-space velocities.
For axion decay and inverse decay, $a\leftrightarrow\gamma\gamma$, the collision term for photons with helicity $\lambda$ has the form
\begin{align}
 C_{\lambda}={}&\frac{1}{2\omega}
 \int\dd\Pi_a\dd\Pi_{\gamma'}(2\pi)^4
 \delta^{(4)}(p_a-k-k')|\mathcal M|^2\notag\\
 &\qquad \times\left[f_a(1+f_\lambda)(1+f_\lambda')
 -f_\lambda f_\lambda'(1+f_a)\right],
 \label{eq:collision-integral}
\end{align}
where $\dd\Pi_j=\dd^3p_j/[(2\pi)^3 2E_j]$, $\mathcal M$ is the decay matrix element, and $p_a$, $k$, and $k'$ are the axion and two photon four-momenta.
The first term describes spontaneous and stimulated axion decay, whereas the second describes inverse decay. 
Here $f_\lambda$ and $f_\lambda'$ are the distribution functions of the two photons, which have the same helicity $\lambda$ in axion decay.

For simplicity, we assume spherical symmetry and that $f_i$ depends only on the radius and energy of species $i$.
At each radius $r$, after integrating over the second photon momentum and the angular variables of the axion momentum, the Boltzmann equation becomes 
\begin{align}
 \frac{\dd f_\lambda(r, \omega,t)}{\dd t}
 ={}&\frac{m_a\Gamma_a}{\omega^2}
 \int_{E_{\min}(\omega)}^\infty\dd E_a\,
 f_a(r, E_a,t)\left[1+f_\lambda(r, \omega,t)
 +f_\lambda(r, E_a-\omega,t)\right]\notag\\
 &-\frac{m_a\Gamma_a}{\omega^2}
 \int_{m_a-\omega}^\infty\dd\omega''\,
 f_\lambda(r, \omega,t)f_\lambda(r,\omega'',t)
 -\frac{f_\lambda(r,\omega,t)}{t_{\rm esc}},
 \label{eq:momentum-boltzmann}
\end{align}
where $\omega=|\bm k|$ and $E_a=(m_a^2+p_a^2)^{1/2}$, with
\begin{equation*}
 E_{\min}(\omega)=\omega+\frac{m_a^2}{4\omega}.
\end{equation*}
Axions with sufficiently large velocities can escape from the local region.
We denote the maximum velocity at position $r$ as $\beta(r)$. 
For the soliton-like halo profile, the characteristic axion momentum is $p_{\rm sol}\simeq1/\rdB$, and we take
\begin{equation}
 \beta(r)=v_{\rm sol}
 =G\Mpbh m_a.
 \label{eq:soliton-velocity}
\end{equation}
In the spike-density region, axions are treated as particles gravitationally bound to the PBH, and we use the local escape velocity,
\begin{equation}
 \beta(r)=v_{\rm esc}(r)=\sqrt{\frac{2G\Mpbh}{r}}.
 \label{eq:spike-velocity}
\end{equation}
The corresponding maximum momentum and energy of a halo axion are
\begin{equation*}
 p_{\max}\simeq m_a\beta,
 \qquad
 E_{\max}=m_a\gamma_\beta,
 \qquad {\rm with}\qquad 
 \gamma_\beta=(1-\beta^2)^{-1/2}.
\end{equation*}
The momentum of a photon produced by axion decay then lies in the range $k_-<\omega<k_+$, with
\begin{equation*}
 k_\pm=\frac{m_a\gamma_\beta}{2}(1\pm\beta).
\end{equation*}
This momentum range follows directly from energy--momentum conservation in the two-body decay $a\to\gamma\gamma$: for an axion moving with speed $\beta$, photon emission parallel and antiparallel to the axion momentum gives $k_+$ and $k_-$, respectively~\cite{Kephart:1994uy}.
In the last line, we added the term $-f_\lambda(\omega,t)/t_{\rm esc}$ to incorporate photon escape from the halo, as discussed further below.

Following the simplified phase-space treatment used in previous studies, we approximate the locally isotropic axion and photon occupation numbers by step-function distributions over their respective momentum intervals~\cite{Kephart:1994uy,Chen:2020momentum}.  
Since the axions are nonrelativistic, $\beta\ll1$, their momenta are restricted to $p\lesssim m_a\beta$, corresponding to a narrow fractional kinetic-energy range $(E_{\max}-m_a)/m_a\simeq\beta^2/2$.  
The decay photons similarly occupy the narrow kinematically allowed interval $k_-<k<k_+$, centered at $m_a\gamma_\beta/2\simeq m_a/2$, with relative half-width $\beta$.  
We therefore neglect the detailed momentum dependence within these intervals and use the corresponding average occupation numbers.  
This approximation preserves the local number densities and the relevant momentum bandwidths while reducing the momentum-dependent Boltzmann equations to local number-density equations
\begin{equation}
 f_a(r,\bm p,t)
 =f_a(r,t)\Theta(m_a\beta-p),
 \qquad
 f_\gamma(r,\bm k,t)
 =f_\gamma(r,t)\Theta(k_+-k)\Theta(k-k_-),
 \label{eq:occupation-average}
\end{equation}
Here, $\Theta$ is the Heaviside step function and $k_\pm$ are defined above.  
The two photon helicities are assumed to be equally populated, $f_+=f_-=f_\gamma/2$, with $f_\gamma=f_++f_-$.
This phase-space averaging reduces the momentum dependent system to local number density equations.

The physical number densities of the axion and photon are
\begin{equation}
 n_a(r,t)=\int\frac{\dd^3p}{(2\pi)^3}f_a(r,\bm p,t),
 \qquad
 n_\gamma(r,t)=\sum_{\lambda=\pm}\int\frac{\dd^3k}{(2\pi)^3}f_\lambda(r,\bm k,t) ,
 \label{eq:number-density-definition}
\end{equation}
where we assume $f_+(\omega)=f_-(\omega)$ for simplicity.
Integrating the collision terms over momentum under these assumptions gives Boltzmann equations for the number density defined in \eq{eq:number-density-definition}
\begin{subequations}
\label{eq:number-density-system}
\begin{align}
 \dot n_\gamma(r,t)={}&2\Gamma_an_a(r,t)
 \left(1+\frac{8\pi^2n_\gamma(r,t)}{m_a^3\beta}\right)
 -\frac{8\pi^2\Gamma_a}{m_a^3}
 \left(1+\frac{2\beta}{3}\right)n_\gamma^2(r,t)
 -\frac{n_\gamma(r,t)}{t_{\rm esc}(r)},\label{eq:number-density-photon}\\
 \dot n_a(r,t)={}&-\Gamma_an_a(r,t)
 \left(1+\frac{8\pi^2n_\gamma(r,t)}{m_a^3\beta}\right)
 +\frac{8\pi^2\Gamma_a\beta}{3m_a^3}n_\gamma^2(r,t)
 +\frac{n_{a,0}(r)-n_a(r,t)}{t_{\rm fill}(r)},\label{eq:number-density-axion}\\
 \dot n_s(r,t)={}&\frac{4\pi^2\Gamma_a}{m_a^3}n_\gamma^2(r,t),
 \label{eq:number-density-sterile}
\end{align}
\end{subequations}
where $n_{a,0}$ is the initial axion number density.

In \eq{eq:number-density-system}, we have added phenomenological terms for photon escape and axion filling.
The photons can escape the coherent de Broglie core with radius $\rdB$ or escape the radius of the local spherical zone for $r>\rdB$, with the escape time given by
\begin{equation}
 t_{\rm esc}(r)=
 \begin{cases}
  2\rdB/(3c), & r\leq\rdB,\\
  2r/(3c), & r>\rdB,
 \end{cases}
 \label{eq:escape-prescription}
\end{equation}
where $c$ is the speed of light.
Photon escape is included in the Boltzmann equation \eq{eq:number-density-photon} through the term
\begin{equation}
 \left.\dot n_\gamma(r,t)\right|_{\rm esc}=-\frac{n_\gamma(r,t)}{t_{\rm esc}(r)}.
 \label{eq:escape-term}
\end{equation}

Axions from the background outside the halo continuously replenish the axion population within the halo. 
This filling term drives the bound axion density toward its undepleted value $n_{a,0}(r)$. 
We represent the filling rate as
\begin{equation}
 \left.\dot n_a(r,t)\right|_{\rm fill}
 =\frac{n_{a,0}(r)-n_a(r,t)}{t_{\rm fill}(r)},
 \label{eq:fill-term}
\end{equation}
where $n_{a,0}(r)$ is the initial number density of local axions and $t_{\rm fill}(r)$ is the filling timescale.  
Since this replenishment is driven by orbital motion of the surrounding bound axions, we characterize it using the gravitational dynamical time,
\begin{equation}
 t_{\rm fill}(r)=
 \begin{cases}
  t_{\rm dyn}(\rdB), & r\leq\rdB,\\
  t_{\rm dyn}(r), & r>\rdB,
 \end{cases}
 \qquad {\rm with} \qquad
 t_{\rm dyn}(r)=\frac{\pi}{2}
 \sqrt{\frac{r^3}{2G\Mpbh}}.
 \label{eq:filling-prescription}
\end{equation}
Within the de Broglie core, the axion configuration is treated as a single coherent region, with the characteristic length scale set to $\rdB$. 
Outside the core, the relevant transport and orbital scales vary with radius, so the local radius $r$ is used instead.

\begin{figure}[t!]
 \centering
 \includegraphics[width=0.82\textwidth]
 {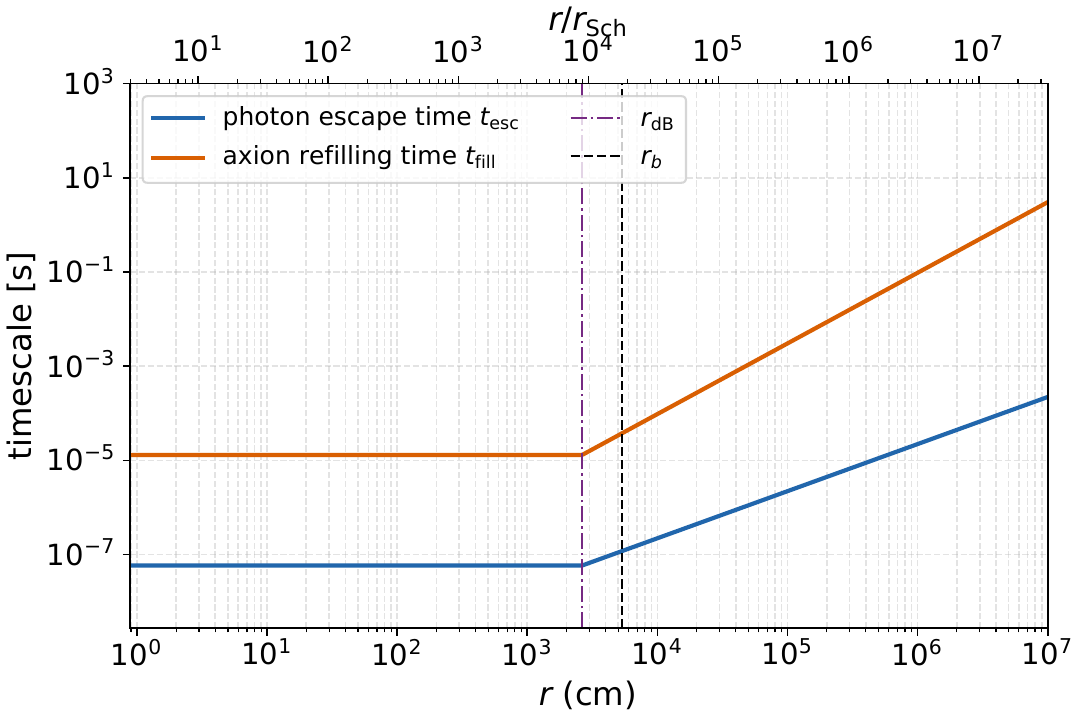}
 \caption{Radial dependence of timescales of photon escape $t_{\rm esc}$ (blue) and axion filling $t_{\rm fill}$ (red). 
 The red dotted, purple dash-dotted, and black dashed vertical lines indicate $r_{\rm cut}$, $r_{\rm dB}$, and $r_b$, respectively.
 Results are shown for the benchmark parameters $M_{\rm PBH}=10^{-6}M_\odot$, $m_a=10^{-6}\,{\rm eV}$, $g_{a\gamma\gamma}= 10^{-11}\,{\rm GeV}^{-1}$, and $\fadm \simeq 1$.}
 \label{fig:transport-timescales}
\end{figure}

In Fig.~\ref{fig:transport-timescales}, we show the radial dependence of the two timescales, $t_{\rm esc}$ and $t_{\rm fill}$, for $M_{\rm PBH}=10^{-6}M_\odot$, $m_a=10^{-6}\,{\rm eV}$, $g_{a\gamma\gamma} = 10^{-11} {\rm GeV}^{-1}$, and $\fadm \simeq 1$.
Both remain constant inside the de Broglie core, whereas outside the core they scale as $t_{\rm esc}\propto r$ and $t_{\rm fill}\propto r^{3/2}$.  
For the reference parameters, $t_{\rm fill}>t_{\rm esc}$ throughout the radial domain, indicating that photons leave the halo faster than the depleted bound axion population is replenished.  
Axion filling can therefore limit the sustained equilibrium emission.  

Inverse decay produces an axion from two photons. 
If the speed of the produced axion exceeds the local escape speed, the axion leaves the halo and does not contribute to the local density of bound axions. 
We denote the number density of these escaping axions by $n_s$. 
Their production term is
\begin{equation}
 \dot n_s(r,t)=\frac{4\pi^2\Gamma_a}{m_a^3}n_\gamma^2(r,t).
 \label{eq:sterile-term}
\end{equation}
The corresponding production term for bound axions is
\begin{equation*}
 \left.\dot n_a(r,t)\right|_{\rm inv,bound}
 =\frac{8\pi^2\Gamma_a\beta}{3m_a^3}n_\gamma^2(r,t),
\end{equation*}
which appears in the axion equation.  

It is useful to define dimensionless variables, normalized by $\tau_a$ and axion mass, given by
\begin{equation}
 \tau=\frac{t}{\tau_a},
 \qquad \bar n_i=\frac{16\pi^2}{m_a^3}n_i,
 \qquad \bar t_{\rm esc}=\frac{t_{\rm esc}}{\tau_a},
 \qquad \bar t_{\rm fill}=\frac{t_{\rm fill}}{\tau_a}.
 \label{eq:dimensionless-variables}
\end{equation}
Then \eq{eq:number-density-system} becomes
\begin{align}
 \frac{\dd\bar n_\gamma}{\dd\tau}
 &=2\bar n_a\left(1+\frac{\bar n_\gamma}{2\beta}\right)
 -\frac12\left(1+\frac{2\beta}{3}\right)\bar n_\gamma^2
 -\frac{\bar n_\gamma}{\bar t_{\rm esc}}, \label{eq:dimless-photon}\\
 \frac{\dd\bar n_a}{\dd\tau}
 &=-\bar n_a\left(1+\frac{\bar n_\gamma}{2\beta}\right)
 +\frac{\beta}{6}\bar n_\gamma^2
 +\frac{\bar n_{a,0}-\bar n_a}{\bar t_{\rm fill}},\label{eq:dimless-axion}\\
 \frac{\dd\bar n_s}{\dd\tau}&=\frac14\bar n_\gamma^2.
\end{align}

\begin{figure}[t!]
 \centering
 \includegraphics[width=0.82\textwidth]{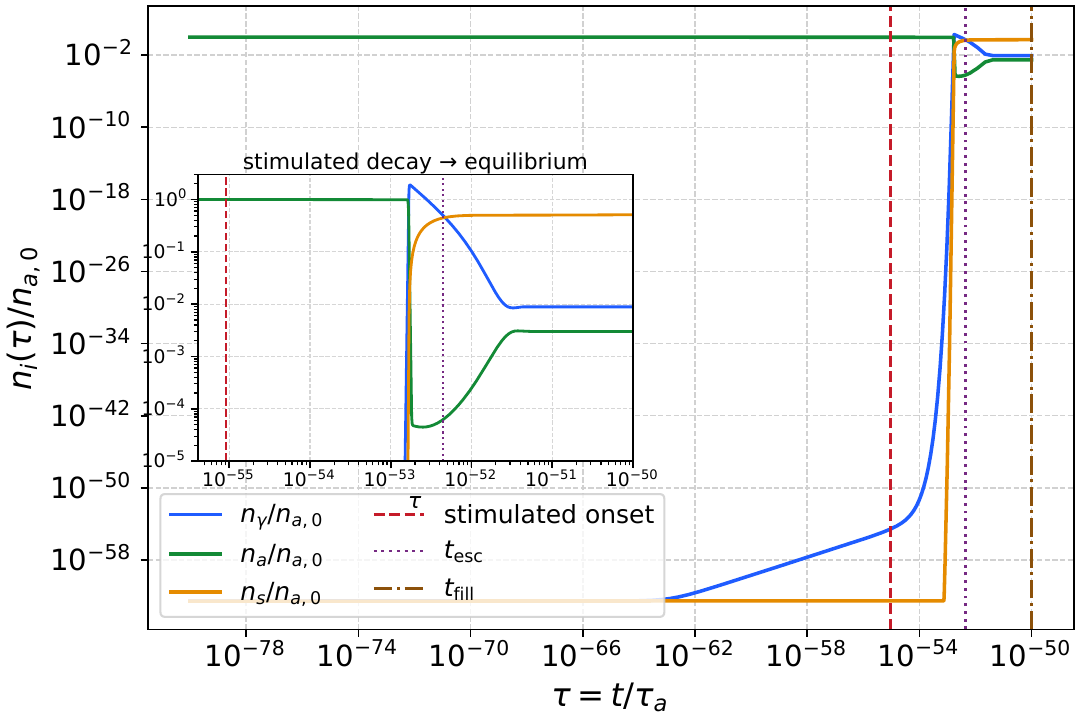}
 \caption{Time evolution of the normalized particle number densities at $r=100\,\rSch $ for the parameters $M_{\rm PBH}=10^{-6}M_\odot$, $m_a=10^{-6}\,{\rm eV}$, $g_{a\gamma\gamma}=10^{-11}\,{\rm GeV}^{-1}$, and $\fadm \simeq 1$.
 The green, blue, and yellow solid curves represent $n_a/n_{a,0}$, $n_\gamma/n_{a,0}$, and $n_s/n_{a,0}$, respectively. 
 The red dashed, purple dotted, and brown dash-dotted vertical lines indicate the estimated onset of stimulated emission $\tau_{\rm onset}$, $\bar t_{\rm esc}$, and $\bar t_{\rm fill}=t_{\rm fill}/\tau_a$, respectively.  The inset enlarges the transition from stimulated decay to equilibrium.
 Here $n_{a,0} \simeq 4.6 \times 10^{32} {\rm cm}^{-3} $, and $\tau_a\simeq 1.3 \times 10^{45}\,{\rm s}$.
 }
 \label{fig:number-density-evolution}
\end{figure}
In Fig.~\ref{fig:number-density-evolution}, we show the time evolution of the axion and photon number densities. 
Here we used the parameters $M_{\rm PBH}=10^{-6}M_\odot$, $m_a=10^{-6}\,{\rm eV}$, $g_{a\gamma\gamma}=10^{-11}\,{\rm GeV}^{-1}$, and $\fadm \simeq 1$.
Initially, spontaneous axion decay seeds the photon population in the halo.
Near the onset time of stimulated emission (vertical red dashed line), the photon number density grows exponentially and depletes the axion population, which is continuously replenished. 
Eventually, both number densities approach their equilibrium values.

%%%%%%%%%%%%%%%%%%%%%%%%%%
\subsection{Local equilibrium and semi-analytic approximations}
%%%%%%%%%%%%%%%%%%%%%%%%%%

At early times, when $\bar n_\gamma/\beta\ll1$ and $\bar n_a\simeq\bar n_{a,0}$, \eq{eq:dimless-photon} can be simplified to
\begin{align}
 \frac{\dd\bar n_\gamma}{\dd\tau}
 &\simeq2\bar n_a\left(1+\frac{\bar n_\gamma}{2\beta}\right),
\end{align}
with solution
\begin{equation}
 \bar n_\gamma(\tau)=2\beta
 \left(e^{\frac{\bar n_{a,0}}{\beta}\tau}-1\right),
\end{equation}
and  $\bar n_\gamma(0)=0$.
Stimulated emission becomes important when $\bar n_{\gamma,{\rm onset}}/2\beta\simeq 1$, and the onset time is therefore
\begin{equation}
 \tau_{\rm onset}=\frac{\beta}{\bar n_{a,0}}.
 \label{eq:onset-time}
\end{equation}

On timescales longer than the photon escape time $t_{\rm esc}$, the system reaches a stationary state in which photon production and loss balance.
In this case, setting $\frac{\dd\bar n_\gamma}{\dd\tau}=0$ in~\eq{eq:dimless-photon} allows us to express the axion number density in terms of the photon number density as
\begin{equation}
 \bar n_a(\bar n_\gamma)=
 \frac{\frac12(1+2\beta/3)\bar n_\gamma^2
 +\bar n_\gamma/\bar t_{\rm esc}}
 {2(1+\bar n_\gamma/(2\beta))}.
 \label{eq:eq-axion-from-photon}
\end{equation}
Substituting this into \eq{eq:dimless-axion} gives
\begin{align}
{}&\frac{1}{4\beta}\bar n_\gamma^3
 +\left[\frac12+\frac{1}{2\bar t_{\rm esc}\beta}
 +\frac{1+2\beta/3}{2\bar t_{\rm fill}}\right]\bar n_\gamma^2\notag\\
&+\left[\frac{1}{\bar t_{\rm esc}}
 \left(1+\frac{1}{\bar t_{\rm fill}}\right)
 -\frac{\bar n_{a,0}}{\beta\bar t_{\rm fill}}\right]\bar n_\gamma
 -\frac{2\bar n_{a,0}}{\bar t_{\rm fill}}=0,
 \label{eq:eq-cubic}
\end{align}
which gives a semi-analytic solution for the equilibrium photon number density.

The condition for the enhancement of stimulated axion decay is $\kappa>0$, where $\kappa$ is defined as
\begin{equation}
 \kappa \equiv \frac{\bar n_{a,0}}{\beta}
 -\frac1{\bar t_{\rm esc}}.
 \label{eq:branch-definitions}
\end{equation}
For a fixed PBH mass, the existence of an enhanced region is determined by the condition $\kappa>0$.
Since $\kappa(r)$ is constant in the core and decreases outward in the spike, its maximum value is $\kappa(\rcut)$. 
Therefore, if $\kappa(\rcut)<0$, then $\kappa(r)<0$ throughout the entire profile.
The condition $\kappa(\rcut)>0$ determines the finite PBH-mass interval in which photon enhancement occurs,
\begin{equation}
 M_{\rm PBH}^{-}<\Mpbh<M_{\rm PBH}^{+},
\end{equation}
where $M_{\rm PBH}^{-}$ and $M_{\rm PBH}^{+}$ denote the lower and upper PBH-mass bounds, respectively. 
The lower bound $M_{\rm PBH}^{-}$ is obtained in the soliton-like regime, whereas the upper bound $M_{\rm PBH}^{+}$ is obtained in the spike-only regime.
Substituting the matched axion profile \eq{eq:matched-axion-profile}, the escape time in \eq{eq:escape-prescription}, and the velocity prescriptions in \eq{eq:soliton-velocity} and \eq{eq:spike-velocity} into the condition $\kappa(\rcut)>0$, and using the dimensionless variables defined in \eq{eq:dimensionless-variables}, we obtain
\begin{equation}
\begin{split}
 M_{\rm PBH}^{-}
 &\simeq
 2.90\times10^{-9}M_\odot
 \left(\frac{\fadm}{1}\right)^{-1}
 \left(
 \frac{\ga}{10^{-11}\,{\rm GeV}^{-1}}
 \right)^{-2}
 \left(
 \frac{m_a}{10^{-6}\,{\rm eV}}
 \right)^{-1/2}, \\
 M_{\rm PBH}^{+}
 &\simeq
 69.6M_\odot
 \left(\frac{\fadm}{1}\right)^2
 \left(
 \frac{\ga}{10^{-11}\,{\rm GeV}^{-1}}
 \right)^4
 \left(
 \frac{m_a}{10^{-6}\,{\rm eV}}
 \right)^{-2}.
 \label{eq:mpbh-enhancement}
 \end{split}
\end{equation}

When stimulated decay becomes dominant, $\bar n_\gamma/(2\beta)\gg1$, and we retain only the stimulated decay contribution in the occupation factor,
\begin{equation*}
 1+\frac{\bar n_\gamma}{2\beta}
 \simeq
 \frac{\bar n_\gamma}{2\beta}.
\end{equation*}
We nevertheless retain the terms for photon escape, axion filling, and inverse decay. 
Under this approximation, the stationary photon equation gives
\begin{equation*}
 \bar n_a\simeq
 \frac{\beta}{\bar t_{\rm esc}}
 +\frac{\beta}{2}
 \left(1+\frac{2\beta}{3}\right)\bar n_\gamma.
\end{equation*}
Substituting this expression into the stationary axion equation, or equivalently using \eq{eq:eq-cubic} in the regime where stimulated decay dominates, gives the following quadratic equation
\begin{equation}
 \bar n_\gamma^2+
 \left[
 \frac{2}{\bar t_{\rm esc}}
 +\frac{2\beta(1+2\beta/3)}{\bar t_{\rm fill}}
 \right]\bar n_\gamma
 -\frac{4\beta\kappa}{\bar t_{\rm fill}}
 =0.
 \label{eq:stim-second-order}
\end{equation}
For $\kappa>0$, the constant term is negative, and hence this quadratic equation has one positive root. 
The positive root is
\begin{equation}
 \bar n_{\gamma,{\rm stim}}
 =\frac{1}{2}
 \left[
 -\left(
 \frac{2}{\bar t_{\rm esc}}
 +\frac{2\beta(1+2\beta/3)}{\bar t_{\rm fill}}
 \right)
 +\sqrt{
 \left(
 \frac{2}{\bar t_{\rm esc}}
 +\frac{2\beta(1+2\beta/3)}{\bar t_{\rm fill}}
 \right)^2
 +\frac{16\beta\kappa}{\bar t_{\rm fill}}
 }
 \right].
 \label{eq:stim-root}
\end{equation}

The condition $\kappa>0$ marks the onset of stimulated enhancement.
For the following limiting forms, we consider the regime in which the net growth rate from stimulated decay is much larger than the photon escape rate,
\begin{equation}
 \kappa\gg\frac{1}{\bar t_{\rm esc}}.
\end{equation}
Using the definition of $\kappa$, we have
\begin{equation}
 \beta\kappa
 =\bar n_{a,0}-\frac{\beta}{\bar t_{\rm esc}}.
\end{equation}
Since the escape contribution is negligible under the above condition, this relation reduces to
\begin{equation}
 \beta\kappa\simeq\bar n_{a,0}.
\end{equation}
Within this regime, the equilibrium photon density depends on the relative importance of photon escape, axion filling, and inverse decay.
\eq{eq:stim-root} therefore has the following three limiting forms.
\begin{itemize}
 \item First, consider the situation in which photons produced by stimulated decay escape from the halo before inverse decay becomes important.
 The photon population is then controlled primarily by the balance between stimulated decay and photon escape, while the depleted axion population is supplied by axion filling. 
 This situation corresponds to
\begin{equation}
 \frac{2}{\bar t_{\rm esc}}
 \gg
 \frac{2\beta(1+2\beta/3)}{\bar t_{\rm fill}},
\end{equation}
and
\begin{equation}
 \frac{16\beta\kappa}{\bar t_{\rm fill}}
 \ll
 \left(\frac{2}{\bar t_{\rm esc}}\right)^2.
\end{equation}
Under these conditions, \eq{eq:stim-root} reduces to
\begin{equation}
 \bar n_{a,{\rm eq}}
 \simeq\frac{\beta}{\bar t_{\rm esc}},
 \qquad
 \bar n_{\gamma,{\rm eq}}
 \simeq
 2\bar n_{a,0}
 \frac{\bar t_{\rm esc}}{\bar t_{\rm fill}}.
 \label{eq:refill-limit}
\end{equation}
 \item Next, consider the situation in which axion filling is sufficiently rapid to maintain the axion density close to its initial value. 
 Stimulated photon production is then balanced mainly by inverse decay, while photon escape is subdominant. 
 The corresponding conditions are
\begin{equation}
 \frac{2\beta(1+2\beta/3)}{\bar t_{\rm fill}}
 \gg
 \frac{2}{\bar t_{\rm esc}},
\end{equation}
and
\begin{equation}
 \frac{16\beta\kappa}{\bar t_{\rm fill}}
 \ll
 \left[
 \frac{2\beta(1+2\beta/3)}{\bar t_{\rm fill}}
 \right]^2.
\end{equation}
\eq{eq:stim-root} then gives
\begin{equation}
 \bar n_{\gamma,{\rm eq}}
 \simeq
 \frac{2\bar n_{a,0}}
 {\beta(1+2\beta/3)}.
 \label{eq:reservoir-limit}
\end{equation}
 \item Finally, consider the situation in which the photon density becomes sufficiently large that inverse decay dominates over the linear contributions associated with photon escape and axion filling. 
 The relevant condition is
\begin{equation}
 \frac{16\beta\kappa}{\bar t_{\rm fill}}
 \gg
 \left[
 \frac{2}{\bar t_{\rm esc}}
 +\frac{2\beta(1+2\beta/3)}{\bar t_{\rm fill}}
 \right]^2.
\end{equation}
In this case, the quadratic inverse-decay contribution dominates, and \eq{eq:stim-root} becomes
\begin{equation}
 \bar n_{\gamma,{\rm eq}}
 \simeq
 2\sqrt{\frac{\beta\kappa}{\bar t_{\rm fill}}}
 \simeq
 2\sqrt{\frac{\bar n_{a,0}}{\bar t_{\rm fill}}}.
 \label{eq:quadratic-limit}
\end{equation}
\end{itemize}

When $\kappa\leq0$, the photon number density cannot be sufficiently enhanced to become comparable to the axion number density before the photons escape from the halo.
Therefore, the axion number density remains equal to its initial value, $\bar n_a=\bar n_{a,0}$, and
\begin{equation}
 \bar n_{\gamma,{\rm spon}}
 =\frac{\kappa+
 \sqrt{\kappa^2+4(1+2\beta/3)\bar n_{a,0}}}
 {1+2\beta/3}.
 \label{eq:subcritical-root}
\end{equation}

In summary, the semi-analytic solution is
\begin{equation}
n_{\gamma,{\rm eq}}^{\rm approx}
 =\frac{m_a^3}{16\pi^2}\bar n_{\gamma,{\rm approx}},
 \quad {\rm with}\quad 
 \bar n_{\gamma,{\rm approx}}=
 \begin{cases}
 \bar n_{\gamma,{\rm spon}},&\kappa\leq0,\\
 \bar n_{\gamma,{\rm stim}},&\kappa>0.
 \end{cases}
 \label{eq:composite-root}
\end{equation}
The corresponding axion density follows directly from \eq{eq:eq-axion-from-photon} for a given $\bar n_\gamma$.

The transition between the two branches in \eq{eq:composite-root} also provides an approximate outer boundary of the stimulated region.
We define $r_{\rm enh}$ by the first radial crossing
\begin{equation}
 \kappa(r_{\rm enh})=0,
 \qquad
 \frac{\bar n_{a,0}(r_{\rm enh})}{\beta(r_{\rm enh})}
 =
 \frac{1}{\Gamma_a t_{\rm esc}(r_{\rm enh})}.
 \label{eq:renh-condition}
\end{equation}
Within the flat-core approximation adopted for the soliton-like region, $\bar n_{a,0}$, $\beta$, and $t_{\rm esc}$ are all independent of $r$.
Consequently, $\kappa$ is also independent of $r$ within the core, and the condition $\kappa=0$ cannot select a particular radius there.
The soliton-like core is therefore either enhanced throughout or non-enhanced throughout. 
A finite enhancement radius arises only in the spike, where
\begin{equation*}
 \bar n_{a,0}(r)\propto r^{-9/4},
 \qquad
 \beta(r)\propto r^{-1/2},
 \qquad
 t_{\rm esc}(r)\propto r.
\end{equation*}
The stimulated contribution $\bar n_{a,0}/\beta\propto r^{-7/4}$ therefore decreases more rapidly than the escape contribution $1/(\Gamma_a t_{\rm esc})\propto r^{-1}$, producing an outer crossing at $r=r_{\rm enh}$.
Using the spike profile in \eq{eq:spike}, together with $\beta(r)=\sqrt{2G\Mpbh/r}$, $t_{\rm esc}(r)=2r/(3c)$, and $r_{\rm eq}=(2G\Mpbh t_{\rm eq}^2)^{1/3}$, we obtain
\begin{equation}
 r_{\rm enh}\simeq
 \left[
 \frac{32\pi^2}{3c}\,
 \fadm\rho_{m,{\rm eq}}t_{\rm eq}^{3/2}
 \Gamma_a(2G\Mpbh)^{1/4}m_a^{-4}
 \right]^{4/3}.
 \label{eq:renh-approx}
\end{equation}
Since $\Gamma_a\propto\ga^2m_a^3$, this result gives
\begin{equation}
 r_{\rm enh}\propto
 \ga^{8/3}\Mpbh^{1/3}m_a^{-4/3}.
\end{equation}
Thus, increasing the axion--photon coupling extends the radial region in which stimulated decay is effective. 
The expression above applies when the crossing occurs in the spike region. 
If the core itself does not satisfy $\kappa>0$, no connected stimulated region extending outward from the central halo is formed.
When \(r_{\rm enh}\leq r_{\rm cut}\), no retained radial interval satisfies \(\kappa(r)>0\). 
We therefore classify this parameter region as having no initial stimulated enhancement.

\begin{figure}[t!]
 \centering
 \includegraphics[width=0.78\textwidth]{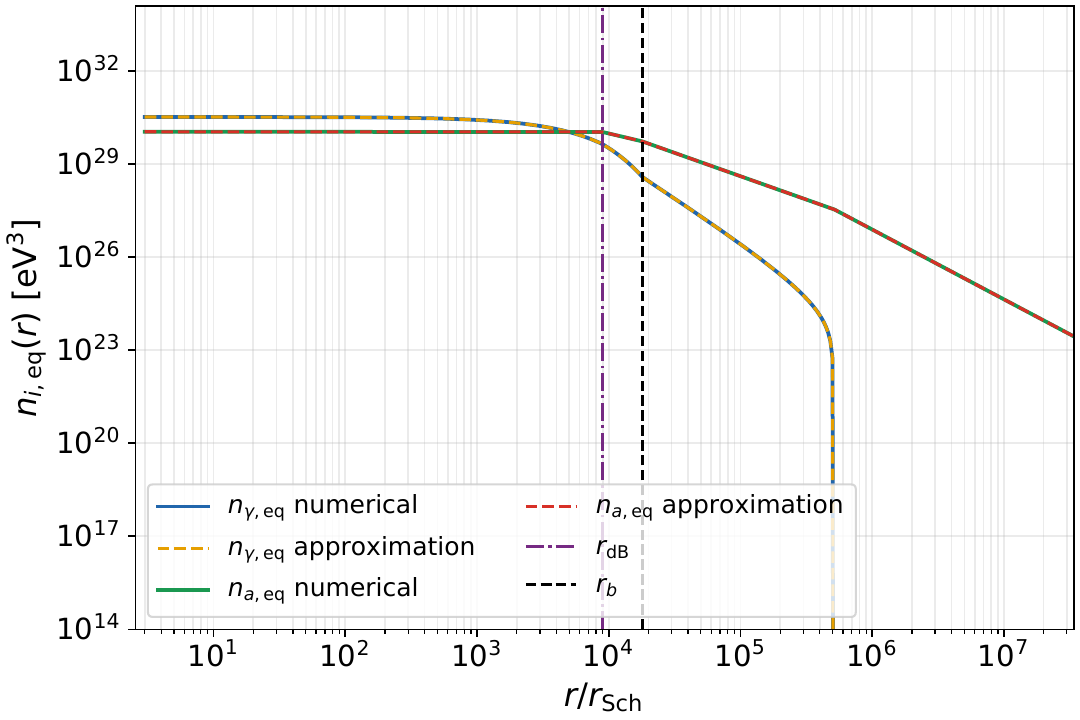}
 \caption{Radial profiles of the equilibrium axion (red) and photon (blue) number densities for the parameters $M_{\rm PBH}=10^{-6}M_\odot$, $m_a=10^{-6}\,{\rm eV}$, $g_{a\gamma\gamma}=10^{-11}\,{\rm GeV}^{-1}$, and $\fadm \simeq 1$.
 The solid and dashed curves represent the numerical and semi-analytic results, respectively. 
 The red dotted, purple dash-dotted, and black dashed vertical lines indicate $r_{\rm cut}$, $r_{\rm dB}$, and $r_b$, respectively.}
 \label{fig:eq-comparison}
\end{figure}

In Fig.~\ref{fig:eq-comparison}, we show the radial profiles of the equilibrium axion (red) and photon (blue) number densities with solid and dashed curves for the numerical results and the semi-analytic results obtained from \eq{eq:stim-root}, respectively, for the parameters $M_{\rm PBH}=10^{-6}M_\odot$, $m_a=10^{-6}\,{\rm eV}$, $g_{a\gamma\gamma}=10^{-11}\,{\rm GeV}^{-1}$, and $\fadm \simeq 1$.
The semi-analytic curves closely match the corresponding numerical results across the entire radial range. 
In the inner stimulated-emission region, axion depletion and filling balance each other. 
At larger radii, stimulated emission becomes inefficient, leading to a rapid decline in the photon density, while the axion density remains at its background value. 
The changes in radial behavior near $r_{\rm dB}$ and $r_b$ reflect, respectively, the change in the photon-escape prescription and the transition from the soliton-like to the spike density profile.
For a detailed view of how the competing Boltzmann terms establish this equilibrium, see Appendix~\ref{app:boltzmann-terms}.

In the analysis above, we assumed that the background axions are supplied by an effectively infinite cosmological reservoir. However, we must also assess the possibility of depleting the cosmological axion population.
The number of background axions within the volume associated with each PBH is given by
\begin{equation}
 N_{a,{\rm bkg}}
 =\frac{1-\fpbh}{\fpbh}\frac{\Mpbh}{m_a}.
 \label{eq:background-reservoir}
\end{equation}
The number of axions used to fill the halo surrounding a PBH for a time duration $\Delta t$ can be estimated as
\begin{equation}
N_{a, \rm fill}
 =
 \int_0^{\Delta t}
 4\pi\int_{\rcut}^{r_{\rm enh}}r^2
 \frac{n_{a,0}(r)-n_{a}(r,t)}{t_{\rm fill}(r)}\dd r  \dd t\leq 
 N_{a, \rm fill}^0=
 \Delta t \times 
 4\pi\int_{\rcut}^{r_{\rm enh}}r^2
 \frac{n_{a,0}(r)}{t_{\rm fill}(r)}\dd r.
 \label{eq:rfill0}
\end{equation}

Then the ratio
\dis{
\Rfillzero \equiv \frac{N_{a, \rm fill}^0}{N_{a,{\rm bkg}}},
\label{eq:axion_filling}
}
quantifies the validity of the assumption that the reservoir of cosmological axions is effectively infinite. 
This assumption is justified when $\Rfillzero\ll 1$.

%%%%%%%%%%%%%%%%%%%%%%%%%%
\section{Diffuse radio background from axion decay in PBH halos}
\label{sec:signal}
%%%%%%%%%%%%%%%%%%%%%%%%%%

Photons escaping from the PBH halo can propagate to Earth, where they may be detected by radio telescopes.
In this section, we calculate the photon spectrum observable at Earth and derive upper bounds on the PBH abundance by comparing our theoretical predictions with observational data.

%%%%%%%%%%%%%%%%%%%%%%%%%%
\subsection{Photon escape rate from one PBH halo}
\label{sec:escape}
%%%%%%%%%%%%%%%%%%%%%%%%%%

At equilibrium, $n_{\gamma,{\rm eq}}/t_{\rm esc}$ in \eq{eq:number-density-photon} is the escaping photon rate per unit physical volume, where $n_{\gamma,{\rm eq}}(r)$ is the local equilibrium photon density and $t_{\rm esc}(r)$ is its escape time. 
A spherical shell of radius $r$ and thickness $\dd r$ contributes to the number of photons escaping from one PBH halo per unit time.
This contribution is given by
\dis{
\frac{\dd R_{\gamma,\rm esc}}{\dd r}(r,t) = 4\pi r^{2}\,\frac{n_{\gamma,\rm eq}(r,t)}{t_{\rm esc}(r)},
}
and the total escape rate of one PBH halo is therefore defined by
\dis{
\qquad
R_{\gamma,\rm esc}(t) \equiv \int_{r_{\rm cut}}^{r_{\rm ta}(t)} \dd r\,
\frac{\dd R_{\gamma,\rm esc}}{\dd r}(r,t),
\label{eq:escape-rate}
}
where $\rta(t)\equiv r_{\rm ta}^{\rm MD}[z(t)]$ is the time-dependent turnaround radius adopted for the assembled halo.
Photon emission is dominated by the enhanced region of the PBH halo and is largely insensitive to $\rta(t)$. 
The detailed cumulative radial contribution is presented in Appendix~\ref{app:radial-source}.

Since the bound axions are nonrelativistic, $\beta\ll1$, photons produced by axion decay have energies near $E_{\rm line}=m_a/2$ and frequencies near $\nu_{\rm line}=E_{\rm line}/h$.
Assuming a monochromatic photon spectrum from each PBH axion halo and accounting for cosmic expansion between emission and observation, we obtain the following observable frequency range
\begin{equation*}
 \left[\frac{\nu_{\rm line}}{1+z_{\rm src,high}},
 \frac{\nu_{\rm line}}{1+z_{\rm src,low}}\right],
\end{equation*}
where $z_{\rm src,high}$ and $z_{\rm src,low}$ denote the initial and final source redshifts, respectively.
The observable frequency range is determined by the overlap between this interval and the telescope's frequency range from 22~MHz to 10~GHz~\cite{Fixsen:2009xn}. 
If there is no overlap, the source cannot be observed.

%%%%%%%%%%%%%%%%%%%%%%%%%%
\subsection{Cosmological diffuse spectrum and observational intensity}
\label{sec:intensity}
%%%%%%%%%%%%%%%%%%%%%%%%%%

To evaluate the cosmological diffuse photon spectrum, we adopt the following simplifying assumptions.
\begin{enumerate}
\item PBHs are statistically homogeneously and isotropically distributed.
\item The redshifted emission contributes to the all-sky diffuse monopole, independent of direction.
\item PBHs have a monochromatic mass distribution, and their comoving number density is conserved.
\end{enumerate}

Under these assumptions, the present-day PBH number density is given by
\begin{equation}
 n_{{\rm PBH},0}=\frac{\fpbh\Omega_{{\rm DM},0}\rho_{c,0}}{\Mpbh},
 \label{eq:pbh-density}
\end{equation}
where $\fpbh$ is the fraction of cosmological dark matter in PBHs, $\Omega_{{\rm DM},0}$ is the present dark matter density parameter and $\rho_{c,0}=3H_0^2/(8\pi G)$ is the present critical density.

Let $E$ be the observed photon energy and $E_{\rm em}=(1+z)E$ its energy of emission at redshift $z$.
The comoving photon-number emissivity per emitted energy is
\dis{
\frac{\dd \dot n_{\gamma,c}}{\dd E_{\rm em}}(E_{\rm em},z)
 = n_{\rm PBH,0}\,\frac{\dd R_{\gamma,\rm esc}}{\dd E_{\rm em}}(E_{\rm em},z),
}
where $\dd\Resc/\dd E_{\rm em}$ is the escaped photon spectrum of one PBH halo, normalized such that $\int\dd E_{\rm em}\,(\dd\Resc/\dd E_{\rm em})=\Resc$, as defined in \eq{eq:escape-rate}.

The present-day photon number density follows by integrating this emissivity over the emission time (or redshift),
\dis{
\frac{\dd n_{\gamma,0}}{\dd E}(E)
&=\int_0^\infty \dd z\,
 \left|\frac{\dd t}{\dd z}\right|
 \frac{\dd E_{\rm em}}{\dd E}
 \frac{\dd\dot n_{\gamma,c}}{\dd E_{\rm em}}
 \bigl((1+z)E,z\bigr)e^{-\tau_\gamma(E,z)}\\
 &= n_{\rm PBH,0}\int_0^\infty \frac{\dd z}{H(z)}\,
 \frac{\dd R_{\gamma,\rm esc}}{\dd E_{\rm em}}\big((1+z)E,\,z\big)\,e^{-\tau_\gamma(E,z)}.
 \label{eq:general-number-density}
}
Here $H(z)$ is the Hubble expansion rate at redshift $z$, $\dd t/\dd z=-[(1+z)H(z)]^{-1}$, and $\dd E_{\rm em}/\dd E=1+z$; the last two factors cancel~\cite{Carr:2009jm,Hogg:1999ad}.  
Equivalently, the physical PBH-density enhancement $(1+z)^3$ is cancelled by the dilution of the freely propagating photon number to the present physical volume.  
The optical depth $\tau_\gamma$ is retained in the general expression and is set to zero in our free-propagation calculation.

For an isotropic radiation field, the photon number intensity per logarithmic observed-energy interval is therefore
\dis{
\mathcal I_N(E) = \frac{c}{4\pi}\frac{\dd n_{\gamma,0}}{\dd \ln E}
 = \frac{c}{4\pi}\,E\,n_{\rm PBH,0}\int_0^\infty \frac{\dd z}{H(z)}\,
 \frac{\dd R_{\gamma,\rm esc}}{\dd E_{\rm em}}\big((1+z)E,\,z\big).
 \label{eq:general-number-intensity}
}
The last equality uses $\tau_\gamma=0$~\cite{Rybicki:1979Radiative}. 
For the monochromatic axion-decay line,
\dis{
\frac{\dd R_{\gamma,\rm esc}}{\dd E_{\rm em}}(E_{\rm em},z)
 = R_{\gamma,\rm esc}(z)\,\delta(E_{\rm em}-E_{\rm line}),
}
where $\delta$ is the Dirac delta distribution, $E_{\rm line}=m_a/2$ is the rest-frame line energy defined above, and $\Resc(z)\equiv\Resc[t(z)]$ from \eq{eq:escape-rate}. 
Therefore, only the redshift $1+z_E=E_{\rm line}/E$ contributes to the spectrum at energy $E$.
Using $\delta[(1+z)E-E_{\rm line}]=\delta(z-z_E)/E$ in \eq{eq:general-number-intensity},
\begin{equation}
 \frac{\dd n_{\gamma,0}}{\dd\ln E}
 =\frac{n_{{\rm PBH},0}\Resc(z_E)}{H(z_E)}.
\end{equation}
Thus, the cosmological integral reduces to
\begin{equation}
 \begin{aligned}
 \IN(E)
 &=\frac{c}{4\pi}
 \frac{n_{{\rm PBH},0}\Resc(z_E)}{H(z_E)}\\
 &\simeq1.24\times10^9\ {\rm m^{-2}\,s^{-1}\,sr^{-1}}
 \frac{H_0}{H(z_E)}
 \left(\frac{\fpbh}{10^{-15}}\right)
 \left(\frac{\Mpbh}{10^{-6}M_\odot}\right)^{-1}\\[-2pt]
 &\hspace{1.8cm}\times
 \left[\frac{\Resc(z_E;\Mpbh,m_a,\ga)}
 {10^{50}\,{\rm s^{-1}}}\right],
 \end{aligned}
 \label{eq:master-intensity}
\end{equation}
where we used $H_0=67.4\,{\rm km\,s^{-1}\,Mpc^{-1}}$, $\Omega_{{\rm DM},0}=0.264$ and $\rho_{c,0}=8.53\times10^{-27}\,{\rm kg\,m^{-3}}$ in the second line~\cite{Planck:2018vyg}. 

The relevant radio observables are~\cite{Rybicki:1979Radiative}
\begin{equation}
 I_\nu=h\IN(E),
 \qquad \nu I_\nu=E\IN(E),
 \qquad T_b=\frac{c^2I_\nu}{2k_B\nu^2},
 \label{eq:radio-observables}
\end{equation}
where $I_\nu$ is the specific intensity at the observed frequency $\nu$, $\nu I_\nu$ is the energy intensity per logarithmic frequency interval, and $T_b$  is the Rayleigh–Jeans brightness temperature. 
Here, $k_B$ denotes Boltzmann's constant.

Using the semi-analytic approximation in \eq{eq:composite-root}, we can estimate $\Resc^{\rm approx}(z)$ as
\begin{equation}
 \Resc^{\rm approx}(z)\simeq\int_{\rcut}^{r_{\rm enh}}
 4\pi r^2\frac{n_{\gamma,{\rm eq}}^{\rm approx}(r)}{t_{\rm esc}(r)}\dd r,
 \label{eq:approx-rate}
\end{equation}
where we assume that $r_{\rm enh}\ll\rta(t_{\rm eq})\leq\rta[t(z)]$, and only include photons produced inside the stimulated region, since emission from this region dominates the photon intensity.
Therefore, the semi-analytic estimate becomes time-independent. 
The corresponding intensity is
\begin{equation}
 \begin{aligned}
 I_\nu^{\rm approx}(\nu)
 &=h\frac{c}{4\pi}
 \frac{n_{{\rm PBH},0}}{H(z_\nu)}\Resc^{\rm approx}(z_\nu)\\
 &\simeq82.0\ {\rm Jy\,sr^{-1}}
 \bfrac{H_0}{H(z_\nu)}
 \left(\frac{\fpbh}{10^{-15}}\right)
 \left(\frac{\Mpbh}{10^{-6}M_\odot}\right)^{-1}\\[-2pt]
 &\hspace{1.8cm}\times
 \left[\frac{\Resc^{\rm approx}(z_\nu;\Mpbh,m_a,\ga)}
 {10^{50}\,{\rm s^{-1}}}\right],
 \end{aligned}
 \label{eq:approx-intensity}
\end{equation}
where $z_\nu$ is selected by $1+z_\nu=m_a/(2h\nu)$, and $1\,{\rm Jy}=10^{-26}\,{\rm W\,m^{-2}\,Hz^{-1}}$.  
Fixed cosmological inputs are absorbed into the coefficient as in \eq{eq:master-intensity}; the remaining dependence is through the free parameters in $\fpbh$, $\Mpbh$, and $\Resc^{\rm approx}(z_\nu;\Mpbh,m_a,\ga)$.

A further simplification is possible in the regime where photon production by stimulated decay is balanced by photon escape. 
In this regime, \eq{eq:refill-limit} gives
\begin{equation}
 n_{\gamma,{\rm eq}}^{\rm approx}(r)
 \simeq
 2n_{a,0}(r)\frac{t_{\rm esc}(r)}{t_{\rm fill}(r)}.
 \label{eq:refill-density-physical}
\end{equation}
Substituting this result into \eq{eq:approx-rate}, we obtain
\begin{equation}
 \Resc^{\rm approx}
 \simeq
 8\pi\int_{r_{\rm cut}}^{r_{\rm enh}} r^2\frac{n_{a,0}(r)}{t_{\rm fill}(r)}\dd r.
 \label{eq:approx-rate-refill}
\end{equation}
The explicit dependence on $t_{\rm esc}$ cancels, showing that the photon emission rate in this regime is regulated by the axion supply into the stimulated region.  
Using the flat-core and spike profiles, $r_b\simeq2r_{\rm dB}$, $r_{\rm cut}=3r_{\rm Sch}$, and $r_{\rm enh}\gg r_b,r_{\rm cut}$, \eq{eq:approx-rate-refill} gives
\begin{equation}
 \Resc^{\rm approx}\simeq
 \begin{cases}
 \displaystyle
 \frac{8}{3}\left(e^4+8\sqrt{2}\right)
 \fadm\rho_{m,{\rm eq}}t_{\rm eq}^{3/2}
 G^2\Mpbh^2m_a^{1/2},
 & \text{hybrid},\\[10pt]
 \displaystyle
 \frac{64\sqrt{2}}{3^{7/4}}
 \fadm\rho_{m,{\rm eq}}t_{\rm eq}^{3/2}
 G^{1/2}\Mpbh^{1/2}m_a^{-1},
 & \text{spike-only}.
 \end{cases}
 \label{eq:approx-rate-branches}
\end{equation}

For $H_0=67.4\,{\rm km\,s^{-1}\,Mpc^{-1}}$, $\Omega_{{\rm DM},0}=0.264$, and the present total matter density parameter $\Omega_{m,0}=0.315$, the matter-era intensity can be written in benchmark-normalized form as
\begin{samepage}
\begin{align}
 I_{\nu,{\rm hybrid}}^{\rm approx}
 &\simeq
 3.98\times10^{6}\,{\rm Jy\,sr^{-1}}
 \left(\frac{\fpbh}{10^{-23}}\right)
 \left(\frac{\fadm}{1}\right)\notag\\[-2pt]
 &\quad\times
 \left(\frac{\Mpbh}{10^{-6}M_\odot}\right)
 \left(\frac{m_a}{10^{-5}\,{\rm eV}}\right)^{-1}
 \left(\frac{\nu}{310\,{\rm MHz}}\right)^{3/2},
 \label{eq:approx-intensity-matched}\\[6pt]
 I_{\nu,{\rm spike\text{-}only}}^{\rm approx}
 &\simeq
 4.62\times10^{4}\,{\rm Jy\,sr^{-1}}
 \left(\frac{\fpbh}{10^{-23}}\right)
 \left(\frac{\fadm}{1}\right)\notag\\[-2pt]
 &\quad\times
 \left(\frac{\Mpbh}{10^{-6}M_\odot}\right)^{-1/2}
 \left(\frac{m_a}{10^{-4}\,{\rm eV}}\right)^{-5/2}
 \left(\frac{\nu}{310\,{\rm MHz}}\right)^{3/2}.
 \label{eq:approx-intensity-spike-only}
\end{align}
\end{samepage}
The two reference masses are chosen so that $310\,{\rm MHz}$ lies within the physical line spectrum of the corresponding branch. 
The leading expressions have no explicit dependence on $\ga$ because it cancels under the approximation that stimulated decay and photon escape are balanced.  
The coupling nevertheless determines whether this regime is reached and fixes $r_{\rm enh}$.

\begin{figure}[t!]
 \centering
 \includegraphics[width=0.65\textwidth]{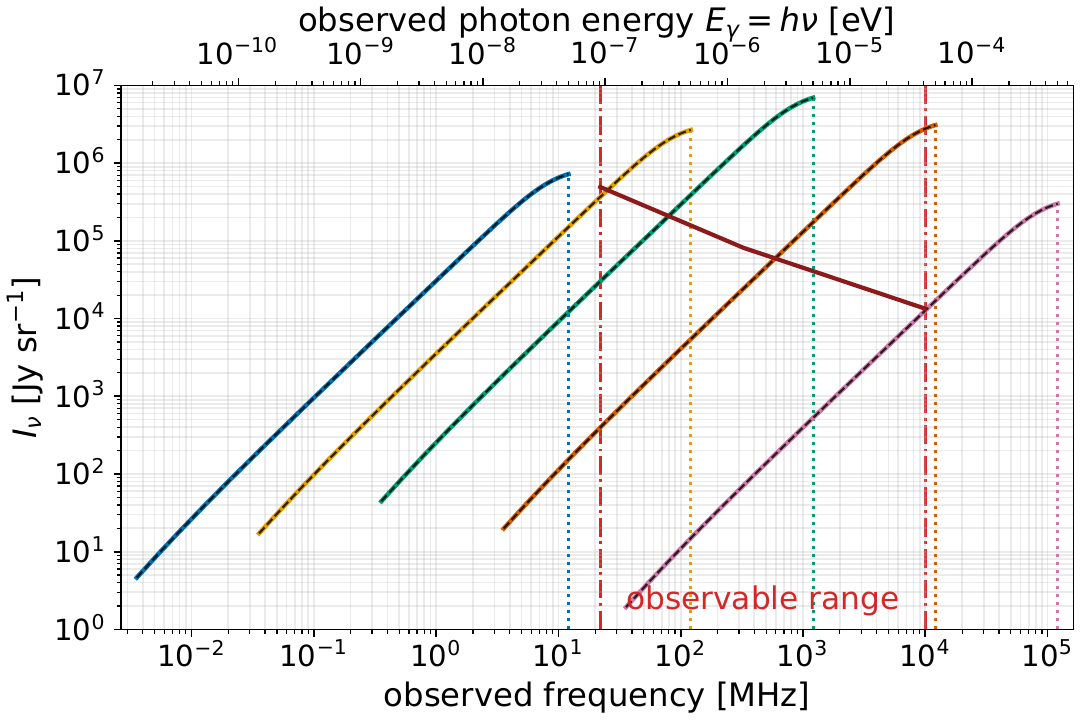}
 \caption{Diffuse photon intensity as a function of the observed frequency for $\Mpbh=10^{-6}M_\odot$, $\ga=10^{-11}\,{\rm GeV}^{-1}$, $\fpbh=10^{-23}$, and $\fadm=1$.  
 The five colored spectra correspond to $m_a=10^{-7},10^{-6},10^{-5},10^{-4}$, and $10^{-3}\,{\rm eV}$.  
 The colored solid curves show the numerical calculation, while the black dashed curves are obtained by inserting the composite equilibrium solution in \eq{eq:composite-root} into \eq{eq:approx-rate} and \eq{eq:approx-intensity}.  
 The upper bound, \eq{eq:radio-envelope} from ARCADE~\cite{Fixsen:2009xn}, is shown by the red line in the observable range.
 The emission redshift range is $0<z<3400$ and the observable frequency range is 22~MHz to 10~GHz~\cite{Fixsen:2009xn}.
 }
 \label{fig:intensity-comparison}
\end{figure}

Fig.~\ref{fig:intensity-comparison} shows the photon intensity obtained from \eq{eq:master-intensity}.
As $m_a$ increases, the curves shift toward higher frequencies, with a cutoff at the maximum frequency $\nu_{\rm line}=m_a/(2h)$. 
For axion masses $m_a<10^{-5}\,{\rm eV}$ in the hybrid regime, the peak intensity increases with mass as $m_a^{1/2}$, while for larger masses in the spike-only regime, the peak intensity decreases as $m_a^{-1}$.
The semi-analytic solutions in \eq{eq:approx-intensity-matched} and \eq{eq:approx-intensity-spike-only} agree very well with the numerical results.

%%%%%%%%%%%%%%%%%%%%%%%%%%
\subsection{Constraint on the PBH abundance from radio observations}
\label{sec:constraint}
%%%%%%%%%%%%%%%%%%%%%%%%%%

Fixsen et al.~\cite{Fixsen:2009xn} combined the direct ARCADE~2 measurements at frequencies above 3~GHz with lower-frequency absolute-temperature surveys and reported the power-law radio excess spectrum
\begin{equation}
 T_b(\nu)=(24.1\pm2.1)\,{\rm K}
 \left(\frac{\nu}{310\,{\rm MHz}}\right)^{-2.599\pm0.036},
 \label{eq:fixsen-fit}
\end{equation}
from 22~MHz to 10~GHz in addition to a CMB temperature.   
We use this combined fit as an upper envelope on an additional radio component,
\begin{equation}
 \begin{aligned}
 I_{\nu,{\rm up}}(\nu)
 &=\frac{2k_B\nu^2}{c^2}T_{b}(\nu)\\
 &=8.36\times10^4\,{\rm Jy\,sr^{-1}}
 \left(\frac{\nu}{310\,{\rm MHz}}\right)^{-0.599},
 \end{aligned}
 \label{eq:radio-envelope}
\end{equation}
where the numerical value is chosen within the $2\sigma$ uncertainty envelope of \eq{eq:fixsen-fit}.

Using the semi-analytic result in \eq{eq:approx-intensity}, we can find the approximate bound at each frequency as
\dis{
\fpbh < & 10^{-15} \bfrac{ I_{\nu,{\rm up}}}{82.0\ {\rm Jy\,sr^{-1}}} 
\bfrac{H_0}{H(z_\nu)}^{-1}
 \left(\frac{\Mpbh}{10^{-6}M_\odot}\right)\\[-2pt]
 &\hspace{1.8cm}\times
 \left[\frac{\Resc(z_\nu;\Mpbh,m_a,\ga)}
 {10^{50}\,{\rm s^{-1}}}\right]^{-1}.
\label{eq:approx-pbh-limit}
}

\begin{figure}[t!]
 \centering
 \includegraphics[width=0.72\textwidth]{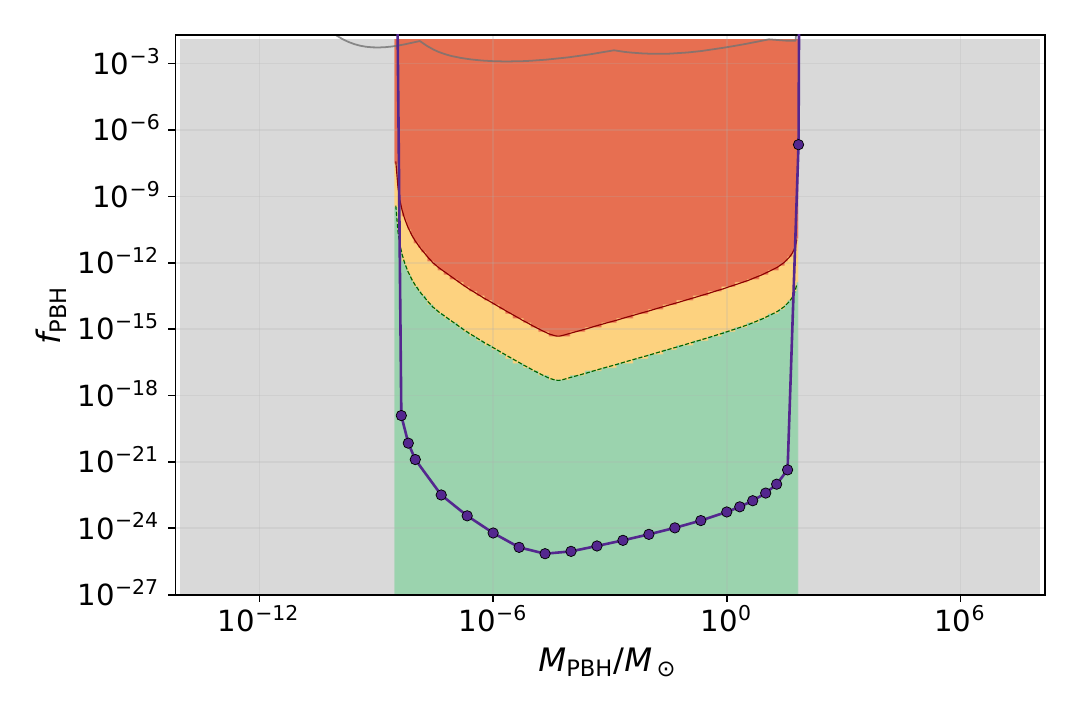}
\caption{Radio-background constraint on the PBH dark-matter fraction $\fpbh$ as a function of $\Mpbh$ for $\ga=10^{-11}\,{\rm GeV}^{-1}$ and $m_a=10^{-6}\,{\rm eV}$.
 Filled purple circles show the numerical results.
 Gray background denotes no initial stimulated enhancement; green, yellow, and red denote enhanced points with $\Rfillzero<10^{-2}$, $10^{-2}\leq\Rfillzero<1$, and $\Rfillzero\geq1$, respectively. 
 Gray curves show general PBH constraints from the \textsc{PBHbounds} compilation for comparison~\cite{Kavanagh:2019PBHbounds}. 
 Curves connecting the sampled masses are visual guides only.
} 
 \label{fig:constraint-single}
\end{figure}

Fig.~\ref{fig:constraint-single} shows the constraints on the PBH dark matter fraction, $\fpbh$, as a function of the PBH mass, $\Mpbh$, for the parameters $\ga=10^{-11}\,{\rm GeV}^{-1}$ and $m_a=10^{-6}\,{\rm eV}$.
In the grey region, the enhancement does not occur ($\kappa(\rcut)<0$). 
In the enhancement region, the background colors show the filling ratio $\Rfillzero$ given in \eq{eq:axion_filling}: $\Rfillzero<10^{-2}$ (green), $10^{-2}\leq\Rfill^{(0)}<1$ (yellow), and $\Rfillzero\geq1$ (red).
Other astrophysical bounds on PBH fraction from the \textsc{PBHbounds} compilation are shown by the gray curves near the top of the figure for comparison~\cite{Kavanagh:2019PBHbounds}. 

Combining \eq{eq:approx-rate-branches} and \eq{eq:approx-pbh-limit}, we can see that the bound on $\fpbh$ scales as $\Mpbh^{-1}$ at small $\Mpbh\lesssim 10^{-5}M_\odot$, where the soliton and spike profiles coexist (the hybrid branch), and as $\Mpbh^{1/2}$ at large $\Mpbh\gtrsim 10^{-5}M_\odot$, where only the spike profile exists (the spike-only branch). 
For the hybrid profile, the axion number density in the soliton core decreases at small $\Mpbh$, resulting in the weaker constraint. 
For the spike-only profile, the radius $\rcut$ increases, which reduces the axion density, so the constraint becomes weaker. 
We checked that the semi-analytic bounds obtained using \eq{eq:approx-pbh-limit} give good agreement with the numerical bounds.

\begin{figure}[h!]
 \centering
 \includegraphics[width=0.95\textwidth]{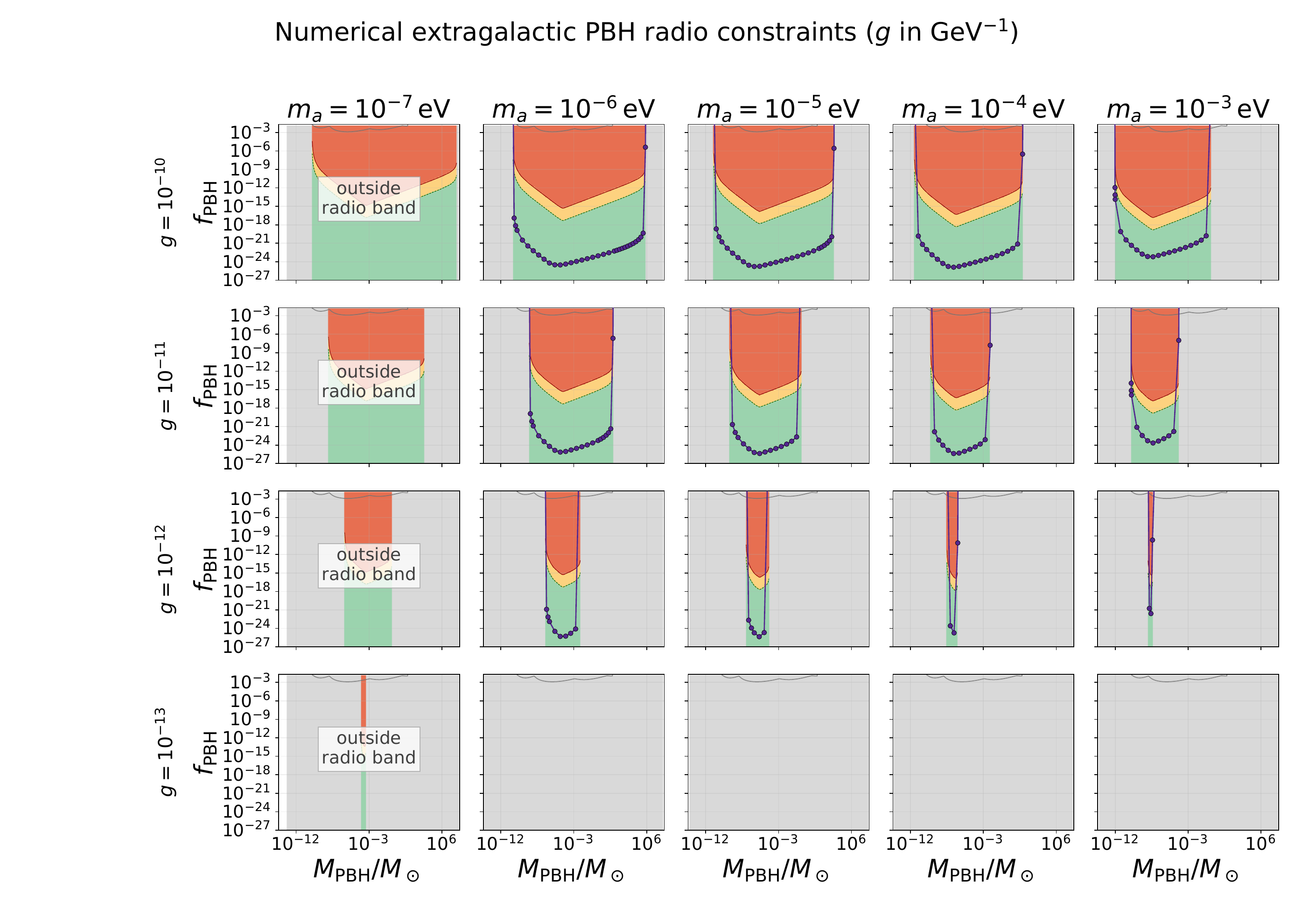}
 \caption{Radio-background constraints on $\fpbh$ over the $(\ga,m_a)$ grid.
 Rows give $\ga$ and columns give $m_a$. 
 Every panel follows the marker, line, and background-color conventions defined in Fig.~\ref{fig:constraint-single}; thus the grid shows the numerical boundary and reservoir-validity diagnostics vary across parameter space.  
 Connecting curves remain visual guides between independently calculated mass points.}
 \label{fig:constraint-grid}
\end{figure}

Fig.~\ref{fig:constraint-grid} extends the calculation over the $(\ga,m_a)$ parameter grid. 
The rows show different values of the axion--photon coupling, while the columns show different axion masses.
The PBH masses at the left and right boundaries are given in \eq{eq:mpbh-enhancement}, with the scaling
\begin{equation}
 \Mpbh^-\propto \ga^{-2}m_a^{-1/2},
 \qquad
 \Mpbh^+\propto \ga^4m_a^{-2}.
\end{equation}
For sufficiently small coupling or sufficiently large axion mass, $M_{\rm PBH}^{-}\geq M_{\rm PBH}^{+}$, so enhancement cannot occur.
As shown in Fig.~\ref{fig:intensity-comparison}, there is no constraint for axion mass $m_a< 2\times 10^{-7}\,{\rm eV}$ since the spectrum is outside the observable radio band.
Once the enhancement occurs, the equilibrium number density does not depend on the coupling constant $\ga$, because the decay-rate dependence cancels out in the equilibrium solution. 

%%%%%%%%%%%%%%%%%%%%%%%%%%
\section{Conclusions}
\label{sec:conclusions}
%%%%%%%%%%%%%%%%%%%%%%%%%%

We have studied stimulated axion decay in the dense axion halos around PBHs, assuming that both PBHs and axions contribute to dark matter, and used the resulting diffuse radio emission to constrain the PBH abundance.

We matched the $r^{-9/4}$ secondary-infall spike to a PBH-dominated hydrogenic soliton-like core outside the inner cutoff $\rcut=3\rSch$, obtaining $\rb\simeq2\rdB$ and a core mass fraction $\epsilon\sim10^{-9}$ for the benchmark parameters. 
The photon--axion Boltzmann equations were reduced to local number-density equations including spontaneous and stimulated decay, inverse decay, photon escape, and axion filling.  
Stimulated enhancement occurs when $\kappa=\bar n_{a,0}/\beta-1/\bar t_{\rm esc}>0$.  
Depending on the relative importance of photon escape, axion filling, and inverse decay, the equilibrium admits three limiting forms. 
In the spike, the stimulated contribution $\bar n_{a,0}/\beta\propto r^{-7/4}$ decreases more rapidly than the escape contribution $1/(\Gamma_a t_{\rm esc})\propto r^{-1}$, confining stimulated emission within $r_{\rm enh}\propto\ga^{8/3}\Mpbh^{1/3}m_a^{-4/3}$.  
Using the CMB photon population as the initial condition would cause stimulated enhancement to begin earlier, but would not change the photon and axion densities reached in dynamical equilibrium.
The semi-analytic solution closely reproduces the numerical equilibrium profiles.

The escaping photon rate is dominated by the stimulated region.  
In the regime where stimulated photon production is balanced by photon escape, it is set by the axion supply and has no explicit dependence on $\ga$. 
It scales as $\Resc\propto\Mpbh^2m_a^{1/2}$ for the hybrid profile and as $\Resc\propto\Mpbh^{1/2}m_a^{-1}$ for the spike-only profile, explaining the rise and fall of the peak intensity with increasing $m_a$.  
Requiring the redshifted diffuse spectrum to lie below the $2\sigma$ ARCADE~2 envelope between 22~MHz and 10~GHz yields upper bounds on $\fpbh$ that are strongest at intermediate PBH masses. 
The constrained PBH-mass interval broadens as $\ga$ increases, while no enhancement occurs for sufficiently small $\ga$.
No radio bound is obtained for $m_a<2\times10^{-7}\,{\rm eV}$, where the line falls below the observed frequency band.  
Once enhancement occurs, the decay rate dependence cancels from the equilibrium density, so the resulting bounds do not follow a simple $\ga^2$ rescaling. 
The semi-analytic bounds closely reproduce the numerical results.
\\

%%%%%%%%%%%%%%%%%%%%%%%%%%
{\it Acknowledgments}.
K.-Y. Choi, and J.Joh acknowledge the financial support from National Research Foundation(NRF) grant funded by the Korea government (MEST) NRF-2022R1A2C1005050.
%%%%%%%%%%%%%%%%%%%%%%%%%%

%%%%%%%%%%%%%%%%%%%%%%%%%%
\appendix

\section*{Appendix}
\addcontentsline{toc}{section}{Appendix}
%%%%%%%%%%%%%%%%%%%%%%%%%%

%%%%%%%%%%%%%%%%%%%%%%%%%%
\section{Evolution of the Boltzmann terms}
\label{app:boltzmann-terms}
%%%%%%%%%%%%%%%%%%%%%%%%%%

\begin{figure}[t!]
 \centering
 \includegraphics[width=0.9\textwidth]
 {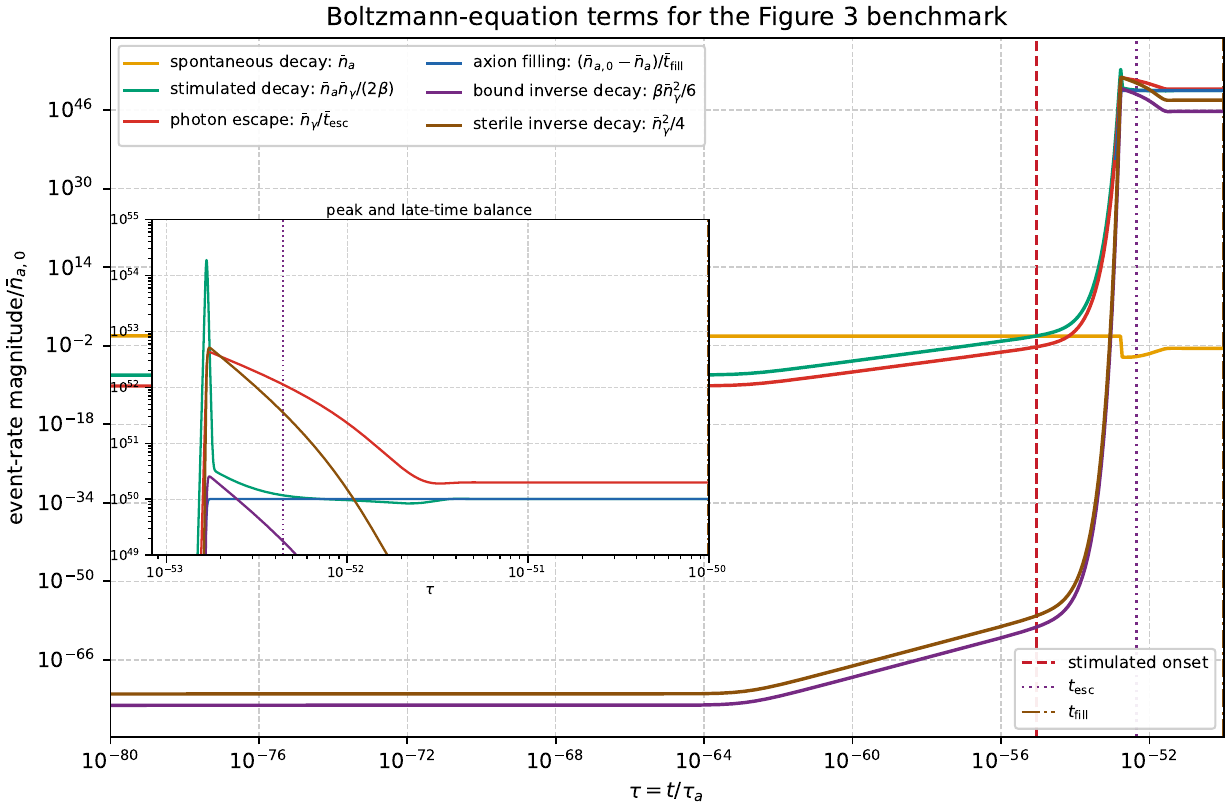}
 \caption{Time evolution of the spontaneous-decay, stimulated-decay,
 photon-escape, axion-filling, and inverse-decay terms for the reference
 benchmark. All terms are normalized by the initial dimensionless axion
 density $\bar n_{a,0}$. The vertical lines indicate the estimated onset of
 stimulated decay and the characteristic escape and filling times.}
 \label{fig:boltzmann-terms}
\end{figure}

To illustrate how the local dynamical equilibrium is established, Fig.~\ref{fig:boltzmann-terms} shows the magnitudes of the individual reaction and transport terms in \eq{eq:number-density-system} for the reference parameters used in Fig.~\ref{fig:number-density-evolution}.
The calculation is evaluated at $r=100\,r_{\rm Sch}$.

At early times, spontaneous decay provides the initial photon population, while the stimulated and inverse-decay terms are negligible.
Once the photon occupation becomes sufficiently large, the stimulated decay term dominates and all terms grow very rapidly.

Finally, the axion number density reaches an equilibrium between stimulated axion depletion and axion filling, while the photon number density reaches an equilibrium between photon production from stimulated axion decay (twice the axion depletion rate) and photon escape.

%%%%%%%%%%%%%%%%%%%%%%%%%%
\section{Radial source and cumulative intensity}
\label{app:radial-source}
%%%%%%%%%%%%%%%%%%%%%%%%%%

\begin{figure}[t!]
    \centering
    \includegraphics[width=0.9\textwidth]
    {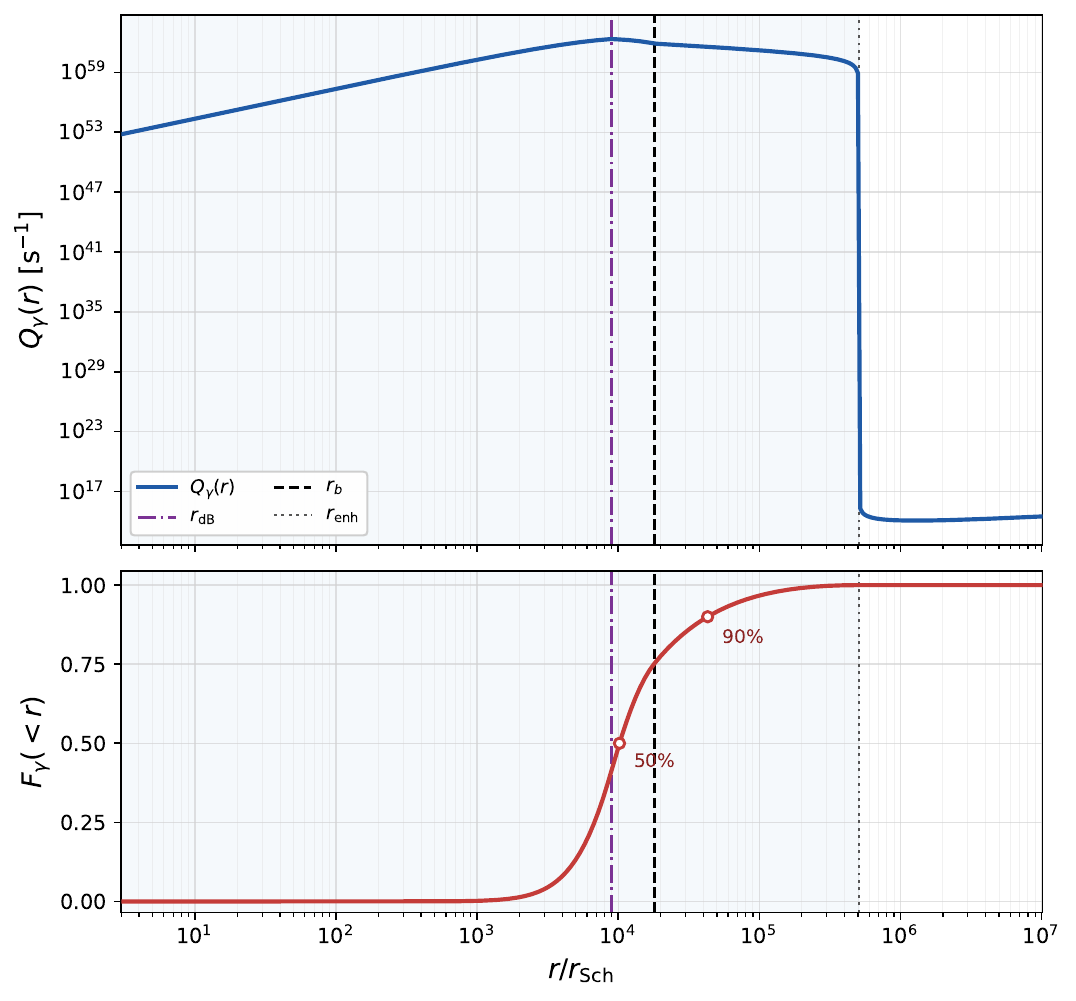}
    \caption{Radial contribution to the escaped photon luminosity, $Q_\gamma(r)$ (upper panel), and its cumulative fraction, $F_\gamma(<r)$ (lower panel), for $M_{\rm PBH}=10^{-6}M_\odot$, $m_a=10^{-6}\,{\rm eV}$, $g_{a\gamma\gamma}=10^{-11}\,{\rm GeV}^{-1}$, and $\fadm=1$.
    The dash-dotted, dashed, and dotted vertical lines indicate $r_{\rm dB}$, $r_b$, and $r_{\rm enh}$, respectively. 
    The open circles mark the radii enclosing $50\%$ and $90\%$ of the total escaped photon luminosity.}
    \label{fig:radial-photon-luminosity}
\end{figure} 

The contribution per logarithmic radius is
\begin{equation}
 Q_\gamma(r)=\frac{\dd R_{\gamma,{\rm esc}}(r)}{\dd\ln r}
 =4\pi r^3\frac{n_{\gamma,{\rm eq}}(r)}{t_{\rm esc}(r)}.
 \label{eq:Qgamma}
\end{equation}
The peak of $Q_\gamma$, rather than the peak of $n_{\gamma,{\rm eq}}$ alone, identifies the radius that dominates the escaped luminosity. 
The cumulative fraction is
\begin{equation}
 F_\gamma(<r)=\frac{1}{\Resc}
 \int_{\rcut}^{r}Q_\gamma(r')\dd\ln r',
\end{equation}
where $\Resc$ is the total escape rate of one PBH halo defined in \eq{eq:escape-rate}.
In Fig.~\ref{fig:radial-photon-luminosity}, we show the radial contribution to the escaped photon luminosity, $Q_\gamma(r)$ (upper panel), and its cumulative fraction, $F_\gamma(<r)$ (lower panel).
We can see that most of the photon emission comes from near the boundary between the soliton and spike regions.

%%%%%%%%%%%%%%%%%%%%%%%%%%
\bibliographystyle{unsrt}
\bibliography{reference}
%%%%%%%%%%%%%%%%%%%%%%%%%%

\end{document}